\documentclass{aa}  

\usepackage{txfonts}
\usepackage[justification=centering]{caption}
 \usepackage{multirow}

\newcommand{\logg}{$\log g$}

\newcommand{\numax}{$\nu_{\mathrm{max}}$}

\newcommand{\dss}{$\delta$~Sct stars}

\newcommand{\filou}{{\sc{filou}}}

\newcommand{\kms}{$\mathrm{km}~\mathrm{s}^{-1}$}

\newcommand{\muhz}{$\mu\mbox{Hz}$}

\newcommand{\Dnulow}{$\Delta\nu_\mathrm{low}$}

\newcommand{\tess}{{\it TESS}}
\newcommand{\multim}{{\sc MultiModes}}
\newcommand{\mesa}{{\sc mesa}}
\newcommand{\myr}{$\mathrm{Myr}$}

\usepackage{indentfirst}
\usepackage[title]{appendix}
\usepackage[usenames]{color}
\usepackage[export]{adjustbox}
\usepackage{float}

\begin{document}

   \title{Dating young open clusters using $\delta$ Scuti stars}

   \subtitle{Results for Trumpler 10 and Praesepe}

   \author{D. Pamos Ortega
          \and
          G. M. Mirouh
          \and
          A. García Hernández
          \and
          J. C. Suárez Yanes
          \and
          S. Barceló Forteza
          }
   \institute{Departamento de Física Teórica y del Cosmos, Universidad de Granada, Campus de Fuentenueva s/n, 18071, Granada, Spain\\
              \email{davidpamos@correo.ugr.es}
             }

   \date{}

 
  \abstract
   {}
   {The main goal of this work is to date young open clusters using \dss. Seismic indices such as the large separation and the frequency at maximum power can help to constrain the models to better characterise the stars. We propose a reliable method to identify some radial modes, which gives us greater confidence in the constrained models.}
   {We extract the frequency content of a sample of  \dss\ belonging to the same open cluster. We estimate the low-order large separation by means of different techniques and the frequency at maximum power for each member of the sample. We use a grid of models built with the typical parameters of \dss, including mass, metallicity and rotation as independent variables, and determine the oscillation modes.  
   We select the observed frequencies whose ratios match those of the models. Once we find a range of radial modes matching the observed frequencies, mainly the fundamental mode, we add it to the other seismic parameters to derive the stellar age. Assuming star groups have similar chemistry and age, we estimate their mean age by computing a weighted probability density function fit to the age distribution of the seismically constrained models.}
   {We estimate the age of Trumpler 10 to be $30_{-20}^{+30}$ \myr, and that of Praesepe to be $580\pm230$ \myr. In this latter case, we find two apparent populations of \dss\ in the same cluster, one at $510\pm140$ \myr\ and another at $890\pm140$ \myr. This may be due to two different formation events, different rotational velocities of the members in our sample of stars (as rapid rotation may modify the observed large separation), or to membership of unresolved binary systems.}
   {}


   \maketitle
%

\section{Introduction}\label{sec:introduction}
   
   Determining the age of a star is essential to know its internal physics. Regarding the dating of a star cluster, the importance lies in understanding the structure and evolution of the galaxy. However, age is not a direct observable and inferring it accurately is not an easy task. In addition, ambiguity arises since we cannot be sure that all the stars in the cluster formed at the same epoch. Recent works have shown that different populations, or generations, of stars may coexist within the same cluster \citep[e.g.][]{Bastian2018}. For instance, \citet{Costa2019} find two distinct populations of stars aged 176 \myr\ and 288 \myr\ in NGC 1866, by combining an analysis of its best-studied Cepheids with that of a very accurate colour-magnitude diagram obtained with the Hubble Space Telescope photometry. Other works, such as \citet{Krause2020}, assume that most open clusters feature a single population, as they remain essentially clear of gas and winds after one stellar formation event. We have assumed this hypothesis in order to determine a mean age for each of thr clusters we analyse.
   
   Traditionally, isochrone fitting on the Hertzsprung-Russell diagram (HRD) has been used to date clusters. This method works when dealing with old globular clusters, where we can find a large sample of stars leaving the main sequence and evolving past the turn-off point. However, the ambiguity of this method is greater with young clusters, in which a majority of stars still on the main sequence (MS). The method based on spectroscopic observations of lithium \citep{Basri1999,Stauffer1999} also generates large ambiguities because of unresolved binary stars \citep{Martin2001}. The relation between the rotation rate and the age of late F to M stars, called gyrochronology \citep{Barnes2003,Angus2022,Messina2022}, seems to provide a method to reduce the uncertainty on the age of evolved clusters. Other methods based on chemical clocks \citep{Silva2012,Spina2018,Moya2022} seem to help reduce the uncertainties, making use of machine learning techniques. The drawback is that these algorithms are trained with models of highly evolved stars, for which it has been possible to obtain reliable spectroscopic observations. Therefore, despite the progress achieved with these new techniques, we still do not have a reliable method to date young open clusters.
   
   In \citet{PamosOrtega} (Paper~I from now on), we proposed the use of seismic parameters to date a group of four \dss\ belonging to the young open cluster $\alpha$ Per. One of these seismic indices is the large separation, defined as the difference between acoustic modes of the same degree and consecutive radial orders, related to the mean density and the surface gravity of the star. This regularity in the frequency pattern is also present in the low-order regime ($n = [2, 8]$) \citep{Suarez2014,GH2015,GH2017,Mirouh2019}, where \dss\ show their oscillation modes. Another parameter is the frequency at maximum power, directly related to the effective temperature, used in solar-type stars and found in \dss\ as well  \citep{Forteza2018,Forteza2020,Bowman2018,Hasanzadeh2021}. 
   
   In this work, we date the young open clusters Trumpler 10 and Praesepe, using a corresponding sample of \dss. We use only seismic parameters, such as the low-order large separation and the frequency at maximum power. We also include the identification of the fundamental mode of each star, estimated by comparing the frequency ratios of the observed frequencies with that of the models. These clusters are of very different ages, which allows us to establish the possibilities of the method. 
   
   The structure of the paper is as follows: in Sect.~\ref{sec:ages}, we provide the estimated ages of the clusters Trumpler 10 and Praesepe from previous works. In Sect.~\ref{sec:data}, we introduce the sample of \dss\ used in this research. In Sect.~\ref{sec:seismic_parameters}, we present how their seismic parameters have been computed. In Sect.~\ref{sec:grid}, we describe the details of the grids of models built to characterise our \dss. In Sect.~\ref{sec:mode_identification}, we explain the method to estimate the mean age of the cluster. In Sect.~\ref{sec:results}, we present the ages we estimate and discuss their reliability by comparing them with the observed parameters and the literature. And finally, in Sect.~\ref{sec:conclusions}, we lay out the main conclusions of our research and leads to improve our method. 

\section{The ages of Trumpler 10 and Praesepe in the literature}\label{sec:ages}

Trumpler 10 and Praesepe have been dated through many different techniques. Table~\ref{tb:Ages_literature} summarises the ages and metallicities derived in the last twenty years. 

Trumpler 10 (C 0846-423) is an open cluster located in the Vela constellation. According to the Milky Way Star Clusters catalogue \citep[MWSC,][]{Kharchenko2013}, it is at a distance of about 417 pc from the Sun and has an age of {$\log({\rm age}) = 7.380$} ($\approx 34$ \myr). According to \citet{Netopil2016}, it has an age of $40\pm10$ \myr, with a metallicity of {$[{\rm Fe}/{\rm H}]$ = $-0.12\pm0.06$}, obtained from various photometric systems and calibrations. The work of \citet{Dias2021} computes the age for this cluster in {$\log({\rm age}) = 7.753\pm0.026$} ($\approx 57$ \myr), with a metallicity of {$[{\rm Fe}/{\rm H}] = 0.043\pm0.050$}. For these estimates, they used Gaia DR2 photometry and a grid of Padova isochrones. According to all these references, the age of Trumpler 10 lies between 34 \myr\ and 57 \myr.

Praesepe (M44, NGC2632) is an open cluster located in the Cancer constellation. Being one of the closest clusters to the Sun, it is also one of the most studied \citep[see for example][and references therein]{Suarez2002, Meibom2005, Fossati2008,Brandt2015,Choi2016,Cummings2017,Gaia2018}. Also taking as reference the MWSC survey, it is at a distance of about 187 pc and has an age of {$\log({\rm age}) = 8.920$} ($\approx 729$ \myr). According to \citet{Netopil2016}, it has an age of $730\pm190$ \myr\ and a metallicity of {$[{\rm Fe}/{\rm H}] = 0.13\pm0.03$}, also obtained from different photometric systems, as for Trumpler 10. The work of \citet{Dias2021} yields {$\log({\rm age}) = 8.882\pm0.035$} ($\approx 762$ \myr), with a metallicity of {$[{\rm Fe}/{\rm H}] = 0.196\pm0.039$}. \citet{Zhong2020} estimate the metallicity at {$[{\rm Fe}/{\rm H}] = 0.22\pm0.08$}, using LAMOST spectroscopy. \citet{Meibom2005} provides an age estimate for Praesepe of about 630 \myr. They used a completely different technique, by looking at the circularization of binary systems of solar-like stars: this circularization happens over time, so systems become circular at higher and higher separations, so that the measurement of the period of circular systems yields an age value. \citet{Douglas2019} estimate an age of $670\pm67$ \myr. Actually, they computed the age of the open cluster Hyades, using a gyrochronology model tuned with slow rotators in Praesepe and the Sun, assuming that the two clusters are coeval, based on the similarity of their colour-magnitude diagrams, activity, rotation and lithium abundance. In short, all these references of the last twenty years provide ages for Praesepe between 590 and 790 \myr.

\begin{table*}
	\centering
	\caption{The ages and metallicities for Trumpler 10 and Praesepe, in some of the main references of the last twenty years, used in this work.}
    \renewcommand{\arraystretch}{1.5}
	\addtolength{\tabcolsep}{2.5 pt}
	\resizebox{15cm}{!}{
    \begin{tabular}{ccccccc}
			\hline
			Reference & Age Trumpler 10 (\myr) & Age Praesepe (\myr) & Z Trumpler 10 & Z Praesepe \\
			\hline
			\citet{Meibom2005} & - & 630 & - & - \\
			\citet{Fossati2008} & - & $590^{+150}_{-120}$ & - & $0.024\pm0.002$  \\
            \citet{Kharchenko2013} & 34 & 729 & - & $0.025\pm0.002$ \\  
            \citet{Netopil2016} & $40\pm10$ & $730\pm190$ & $0.014\pm0.002$ & $0.024\pm0.002$ \\
            \citet{Douglas2019} & - & $670\pm67$ & - & - \\
            \citet{Zhong2020} & - & - & - & $0.031\pm0.005$ \\
            \citet{Dias2021} & 57 & 762 & $0.018\pm0.004$ & $0.028\pm0.002$ \\
			
    \end{tabular}
    }
	\label{tb:Ages_literature}
\end{table*}

\section{The data}\label{sec:data}

Firstly, we cross-match the VizieR Online Data Catalogue Gaia DR2 of open cluster members \citep{Cantat-Gaudin2018} and the TESS Input Catalogue \citep[TIC,][]{Stassun2019}, searching possible \dss\  belonging to the same open cluster. 

According to the definition of a pure $\delta$ Sct from \citet{Griga2010} and \citet{Uytterhoeven2011}, we find five candidates in the field of Trumpler 10 (Table~\ref{tb:parametersTr10}) and six in the field of Praesepe (Table~\ref{tb:parametersM44}). For our Praesepe stars, we find values for the projected rotational velocity $v\sin i$, the metallicity and the spectral type, consulting the available references in the Simbad Astronomy Database\footnote{https://simbad.unistra.fr/simbad/}. For the Trumpler 10 stars, only data about the spectral type is available. These parameters are useful in our discussion of the results in Sect.~\ref{sec:results}.

We perform a frequency analysis using data from sector 35 of the \tess\ mission \citep{Ricker} for the Trumpler 10 sample, with approximately 13800 points; and from sector 45 for Praesepe, with approximately 15500 points. In both cases, the cadence is about 2 minutes and the Rayleigh resolution is approximately 0.041 $d^{-1}$. We use the Pre-Search Data Conditioned (PDC) light curves, corrected for instrumental effects, that are publicly available through the TESS Asteroseismic Science Consortium\footnote{https://tasoc.dk} (TASC).

Using \multim\footnote{https://github.com/davidpamos/MultiModes}, we extract the frequency content of each star in our sample. It is a Python code that calculates the Fast Lomb-Scargle periodogram \citep{Press} of a light curve. It extracts, one by one, a limited number of significant signals, and uses their corresponding parameters (amplitudes, frequencies) to fit the total signal to a multisine function with a non-linear least squares minimisation. We adopt a signal-to-noise ratio S/N > 4.0 as a stop criterion \citep{Breger1993}, to avoid spurious frequencies, and we filter possible frequency combinations. The code is presented in detail in Paper~I and in the public repository.

\begin{figure*}   
    \centering
    \includegraphics[width=18cm,height=10cm,keepaspectratio]{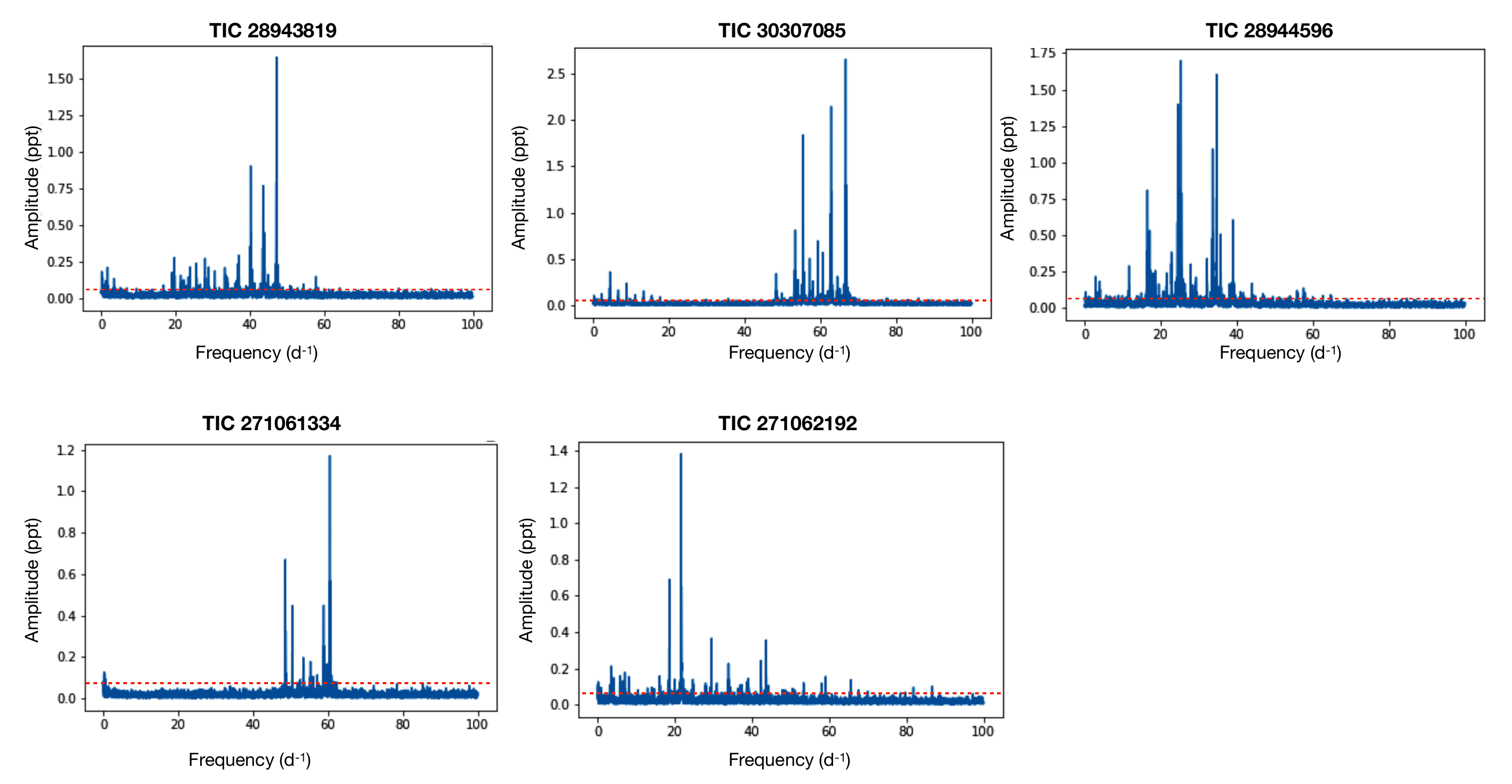}
    \caption{Frequency spectrum of the sample of \dss\ candidates in the field of Trumpler 10. The red dotted line is the significance threshold, S/N=4.}
    \label{fig:Spectrum_Trumpler10}
\end{figure*}

\begin{figure*}   
    \centering
    \includegraphics[width=18cm,height=10cm,keepaspectratio]{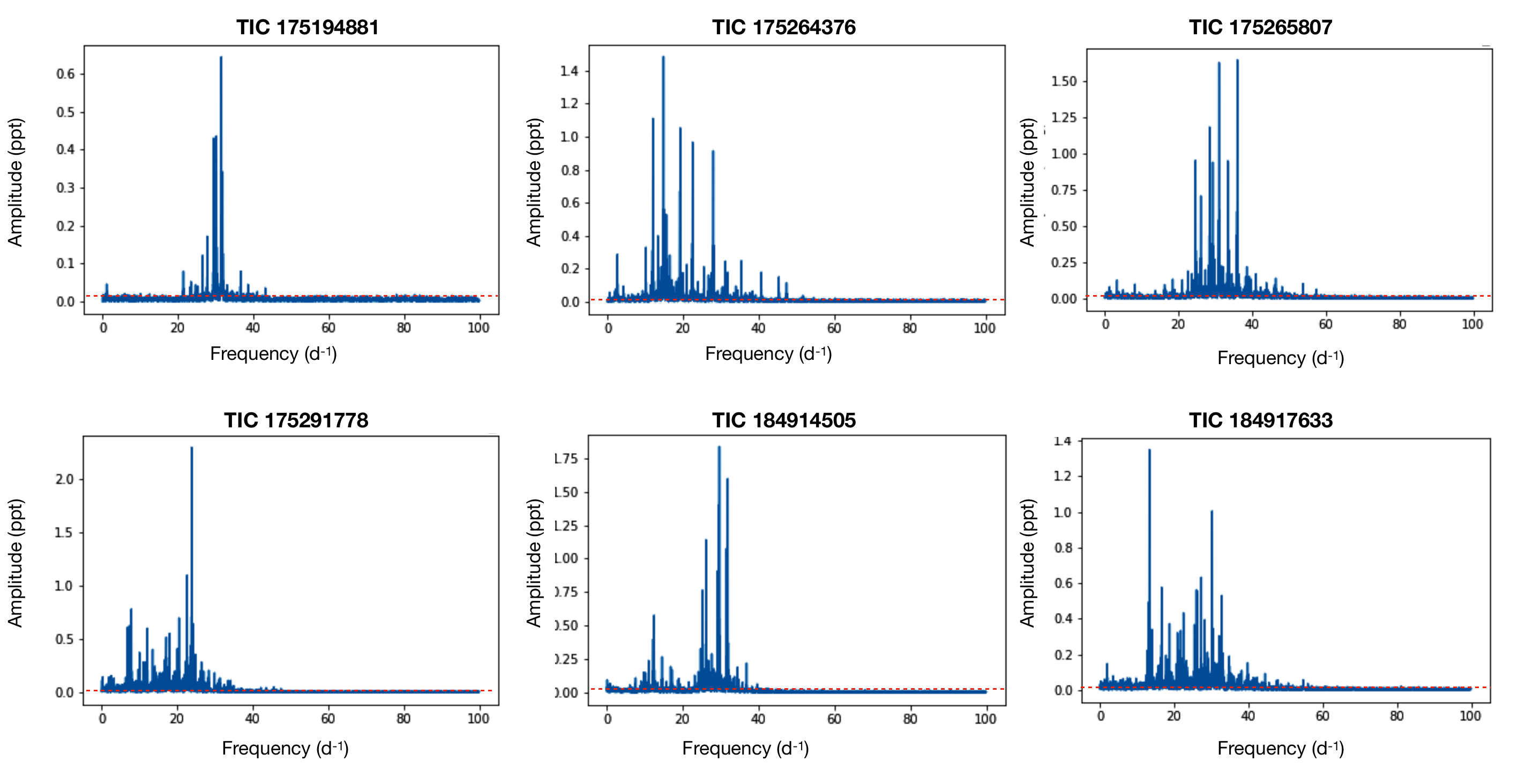}
    \caption{Same as Fig.~\ref{fig:Spectrum_Trumpler10} for the Praesepe sample.}
    \label{fig:Spectrum_Praesepe}
\end{figure*}

Fig.~\ref{fig:Spectrum_Trumpler10} and Fig.~\ref{fig:Spectrum_Praesepe} show, respectively, the extracted frequency spectrum of each $\delta$ Sct candidate in Trumpler 10 and Praesepe. The values of the ten highest amplitudes frequencies for each star in both clusters are shown in Table~\ref{tb:frequencies} (Appendix~\ref{appendixA}). The table of all extracted frequencies is available on-line. 

\begin{table*}
	\centering
	\caption{Stellar parameters of the selected targets in Trumpler 10. From left to right: TESS magnitude, luminosity, mass, radii, density, surface gravity, effective temperature, cluster member probability and spectral type. References: $^{(1)}$\citet{Stassun2019},
	$^{(2)}$\citet{Cantat-Gaudin2018},
    $^{(3)}$\citet{Skiff2014}.}
    \renewcommand{\arraystretch}{1.5}
	\addtolength{\tabcolsep}{2.5 pt}
	\resizebox{18cm}{!}{
    \begin{tabular}{cccccccccc}
			\hline
			TIC & $T_{mag}^{(1)}$ & $\log (L/L_{\odot})^{(1)}$ & $M (M_{\odot})^{(1)}$ & $R (R_{\odot})^{(1)}$ & $\bar \rho (\bar \rho_{\odot})$ & $\log g^{(1)}$ & $T_\mathrm{eff}(K)^{(1)}$ & $P_{member}^{(2)}$ &
            Spectral type$^{(3)}$\\
			\hline
			28943819 & $10.545\pm0.007$ & $1.11\pm0.04$ & $2.2\pm0.3$ & $1.60\pm0.06$ & $0.53\pm0.14$ & $4.37\pm0.08$ & $8646\pm161$ & 1.0 & - \\
			30307085 & $10.294\pm0.006$ & $1.28\pm0.04$ & $2.5\pm0.3$ & $1.47\pm0.04$ & $0.80\pm0.17$ & $4.51\pm0.06$ & $9931\pm202$ & 0.8 & A0  \\
			28944596 & $10.354\pm0.006$ & $1.12\pm0.03$ & $2.1\pm0.3$ & $1.72\pm0.06$ & $0.41\pm0.10$ & $4.28\pm0.07$ & $8383\pm149$ & 1.0 & A2 \\ 
			271061334 & $10.347\pm0.006$ & $1.12\pm0.03$ & $2.2\pm0.3$ & $1.58\pm0.05$ & $0.56\pm0.13$ & $4.38\pm0.07$ & $8773\pm170$ & 0.9 & - \\
			271062192 & $10.238\pm0.006$ & $1.16\pm0.03$ & $2.2\pm0.3$ & $1.68\pm0.05$ & $0.46\pm0.11$ & $4.33\pm0.07$ & $8689\pm158$ & 1.0 & A3 \\
    \end{tabular}
    }
	\label{tb:parametersTr10}
\end{table*}

\begin{table*}
	\centering
	\caption{Stellar parameters of the selected targets in Praesepe. From left to right: TESS magnitude, luminosity, mass, radii, density, surface gravity, effective temperature, cluster member probability, projected rotational velocity, metallicity and spectral type. References: $^{(1)}$\citet{Stassun2019},
	$^{(2)}$\citet{Cantat-Gaudin2018},
    $^{(3)}$\citet{Cummings2018},
    $^{(4)}$\citet{Fossati2008},
    $^{(5)}$\citet{Bochanski2018}.}
    \renewcommand{\arraystretch}{1.7}
	\addtolength{\tabcolsep}{2.5 pt}
	\resizebox{18cm}{!}{
    \begin{tabular}{ccccccccccccc}
			\hline
			TIC & $T_{mag}^{(1)}$ & $\log (L/L_{\odot})^{(1)}$ & $M (M_{\odot})^{(1)}$ & $R (R_{\odot})^{(1)}$ & $\bar \rho (\bar \rho_{\odot})$ & $\log g^{(1)}$ & $T_\mathrm{eff}(K)^{(1)}$ & $P_{member}^{(2)}$ & v sini (km/s) & [Fe/H]$^{(4)}$ & [Fe/H]$^{(5)}$ & Spectral type$^{(4)}$  \\
			\hline
			175194881 & $7.966\pm0.006$ & $1.18\pm0.02$ & $1.9\pm0.3$ & $2.10\pm0.07$ & $0.20\pm0.05$ & $4.07\pm0.08$ & $7873\pm125$ & 1.0 & $85^{(4)}$ &
            0.26 & - & A7V  \\
			175264376 & $8.233\pm0.008$ & $1.09\pm0.01$ & $1.7\pm0.3$ & $2.13\pm0.09$ & $0.17\pm0.05$ & $4.01\pm0.08$ & $7416\pm141$ & 1.0 & $200^{(4)}$ &
            0.12 & 0.08 & F0Vn\\
			175265807 & $8.134\pm0.006$ & $1.10\pm0.02$ & $1.9\pm0.3$ & $1.92\pm0.06$ & $0.26\pm0.07$ & $4.14\pm0.08$ & $7826\pm126$ & 1.0  & $135^{(3)}$ &
            - & 0.01 & - \\
			175291778 & $7.771\pm0.006$ & $1.26\pm0.01$ & $1.9\pm0.3$ & $2.31\pm0.06$ & $0.15\pm0.04$ & $3.98\pm0.08$ & $7865\pm126$ & 1.0 & $150^{(4)}$ &
            0.09 & -0.08 & A7V \\
			184914505 & $8.255\pm0.007$ & $1.00\pm0.02$ & $1.7\pm0.3$ & $1.94\pm0.06$ & $0.23\pm0.06$ & $4.09\pm0.08$ & $7369\pm108$ & 1.0 & $90^{(4)}$ &
            0.31 & - & A5\\
			184917633 & $8.155\pm0.006$ & $1.08\pm0.02$ & $1.7\pm0.3$ & $2.08\pm0.07$ & $0.19\pm0.05$ & $4.03\pm0.08$ & $7443\pm122$ & 1.0 & $155^{(4)}$ &
            -0.02 & -0.02 & A5 \\
    \end{tabular}
    }
	\label{tb:parametersM44}
\end{table*}

\section{Seismic parameters}\label{sec:seismic_parameters}

\begin{figure*}   
    \centering
    \includegraphics[width=0.8\textwidth]{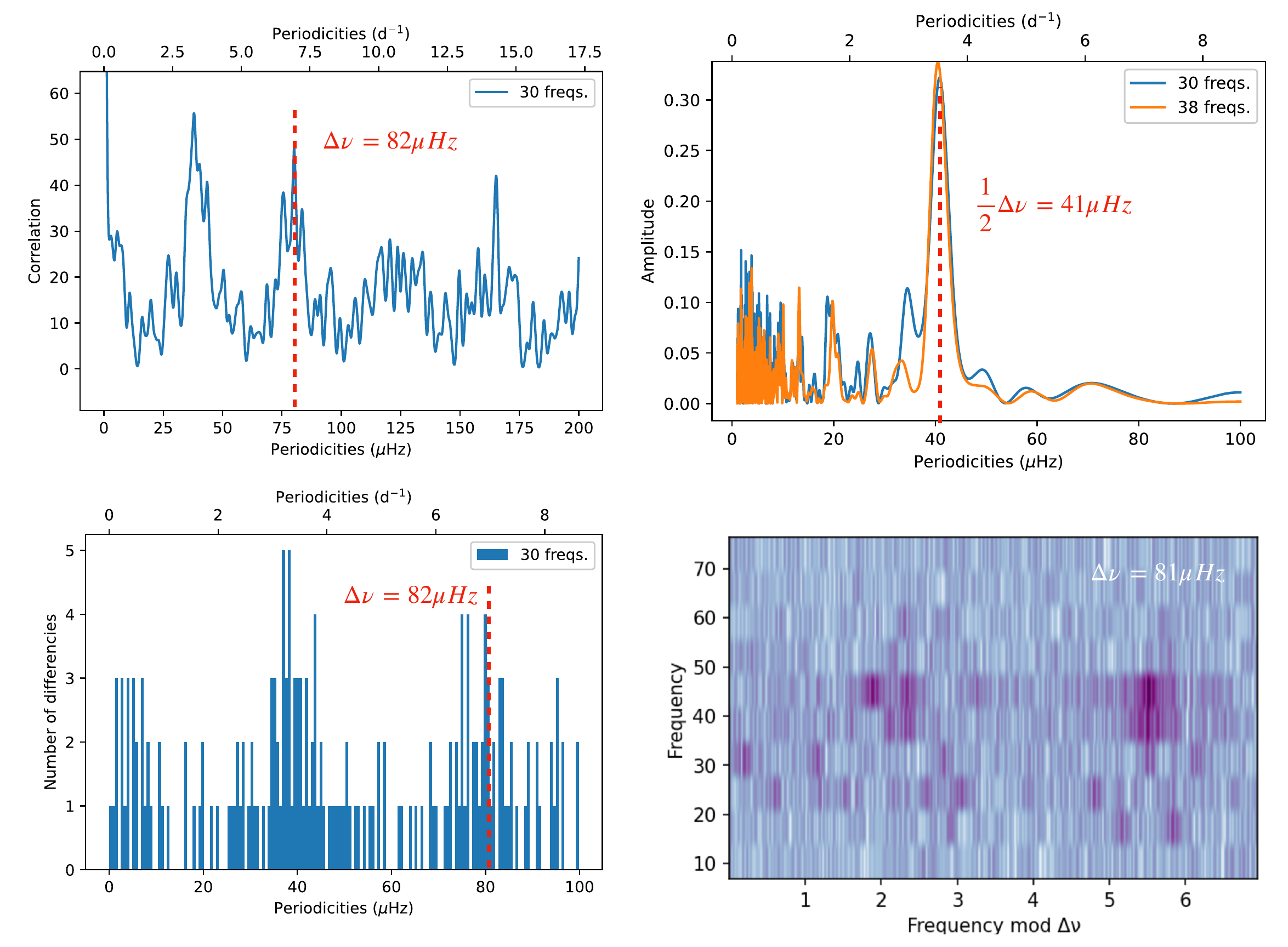}
    \caption{Estimated low-order large separation for TIC\,28943819, using the autocorrelation diagram (top left), the discrete Fourier transform (top right), the frequency difference histogram (bottom left), and the échelle diagram \citep[][bottom right]{Hey2020}.}
    \label{fig:TIC28943819_large_separation}
\end{figure*}

Sometimes it is possible to find regularities in the complex frequency pattern of a $\delta$ Sct star \citep{GH2009,Paparo2016,Bedding2020}. Following the same techniques as in \citet{GH2009, GH2013, RB2021} and used in Paper~I, we estimate the large separation in the low-order regime, $\Delta\nu_{low}$. Fig.~\ref{fig:TIC28943819_large_separation} shows an example of this, where we use the discrete Fourier transform (DFT), the autocorrelation diagram (AC) applied on the frequencies, the frequency difference histogram (FDH) and the échelle diagram (ED), in order to find regularities in the frequency content of TIC\,28943819 (see Appendix~\ref{appendixB} for the rest of our sample). The theoretical works of \citet{GH2009,Reese2017} use the AC and the FT to search for the low-order large separation. They point out that we expect to see the large separation and its submultiples in the DFT, and its multiples in the AC and the FDH (except in the case where the $l=1$ modes lie halfway between the $l=0$ modes, as in solar-type stars, where we also find half the large separation). No method, by itself, is objective enough to yield a reliable measurement of $\Delta\nu_{low}$, except in very few cases. Our criterion requires finding the same peaks in at least two of the methods used for measuring \Dnulow~or half \Dnulow. We estimate the uncertainties using the width of the peaks in the DFT or the AC, depending on the case. Of all the stars analysed this way, TIC\,28944596 and TIC\,271062192 are the most difficult cases. TIC\,28944596 (see Fig.~\ref{fig:28944596_large_separation}) only has a few frequencies. Its DFT shows a peak a bit above 20 \muhz\ and another around 40 \muhz. The AC also shows two peaks very close to 80 \muhz, and the FDH shows three peaks between 70 and 80 \muhz. For all these reasons, we estimate the large separation to be 80 \muhz\ for this star. TIC\,271062192 (Fig.~\ref{fig:271062192_large_separation}) is an even more complicated case, because the AC, the FDH and the ED do not show regularities in the frequency content. The only evidence here is the DFT, that shows three congruent peaks around 19, 38 and 76 \muhz. This is why we have retained a large separation estimate of 76 \muhz. 

\citet{Forteza2020} find a scaling relation between the frequency at maximum power (\numax) and the effective temperature in a sample of more than 2000 \dss. They define \numax\ as the average frequency weighted over the amplitudes, $A_{i}$, of the frequencies, $\nu_{i}$, of the envelope,
\begin{equation}\label{eq1}
    \nu_{\rm max} = \frac{\sum{(A_{i}\cdot\nu_{i})}}{\sum{A_{i}}}.
\end{equation}

Another independent study \citep{Hasanzadeh2021} finds a similar correlation with a sample of more than 400 \dss. They use the AC method on several windows of the S/N periodogram, to obtain the range of frequencies for the modes' envelope. Then, a Gaussian curve is fitted to the mean collapsed correlation with each window. The peak of this curve is their \numax. However, such correlation between \numax\ and the effective temperature is recently debated by \citet{Bedding2023}, who do not see it with a sample of 35 \dss\ in the Pleiades cluster. The \numax\ definitions of \citet{Forteza2020} and \citet{Hasanzadeh2021}, both based on the envelope, are more stable than choosing the frequency with the highest amplitude, as is done by \citet{Bedding2023}, since this frequency is more likely to vary with time in the presence of non-linear effects than the envelope. Despite the large uncertainty involved in using \numax\ as a seismic parameter, we decide to trust it as a seismic indicator. 

Following \citet{Forteza2020}, we relate $\nu_{\rm max}$ and $\tilde{T}_\mathrm{eff}$. The relation depends on the value of $\log g$, so for Trumpler 10 ($\log g \approx4.3)$ we use
\begin{equation}\label{eq2}
    \Tilde{T}_\mathrm{eff} = (3.5\pm0.1)\nu_{\rm max}^{(\mu Hz)} + (6460\pm40)^{(K)} ,
\end{equation}
while for Praesepe ($\log g \approx4.0)$ we use
\begin{equation}\label{eq3}
    \Tilde{T}_\mathrm{eff} = (3.8\pm0.2)\nu_{\rm max}^{(\mu Hz)} + (6750\pm40)^{(K)}.
\end{equation}


The estimated values for $\Delta\nu_{\rm low}$, \numax\ and their corresponding seismic temperatures, $\tilde{T}_\mathrm{eff}$, for our sample of \dss\ in Trumpler 10 and Praesepe are shown in Table~\ref{tb:indices}. 

Comparing the values of TIC effective temperature and the seismic temperature, we see that the agreement is good for the majority of the sample, taking into account the large uncertainties of the seismic temperature. However, TIC\,30307085 and TIC\,271062192 are discrepant, with differences between the two estimates above 1000K, which can be the signature of a binary companion or gravity darkening induced by rapid rotation. 

\begin{table}
	\centering
	\caption{Seismic indices of the selected targets from Trumpler 10 and Praesepe: the low-order large separation, the frequency at maximum power and its corresponding seismic and TIC effective temperature.}
    \renewcommand{\arraystretch}{1.3}
	\addtolength{\tabcolsep}{2.0 pt}
	\resizebox{7cm}{!}{
    \begin{tabular}{ccccc}
			\hline
			TIC & \Dnulow (\muhz) & \numax (\muhz) & $\tilde{T}_\mathrm{eff}$ (K) & TIC $T_\mathrm{eff}$ (K)\\
			\hline
            \multicolumn{4}{c}{Trumpler 10}\\
            \hline
			28943819 & $82\pm2$ & $510\pm30$ & $8250\pm200$ & $8646\pm161$\\
			30307085 & $84\pm1$ & $710\pm60$ & $8950\pm320$ & $9931\pm202$\\
			28944596 & $80\pm2$ & $330\pm80$ & $7620\pm350$ & $8383\pm149$\\
			271061334 & $80\pm2$ & $650\pm50$ & $8740\pm280$ & $8773\pm170$\\
			271062192 & $76\pm2$ & $290\pm90$ & $7480\pm380$ & $8689\pm158$\\
            \hline
			\multicolumn{4}{c}{Praesepe}\\
            \hline
			175194881 & $58\pm1$ & $350\pm30$ & $8080\pm220$ & $7873\pm125$\\
			175264376 & $52\pm3$ & $210\pm60$ & $7550\pm310$ & $7416\pm141$\\
			175265807 & $57\pm2$ & $360\pm40$ & $8120\pm260$ & $7826\pm126$\\
			175291778 & $52\pm3$ & $200\pm70$ & $7510\pm350$ & $7865\pm126$\\
			184914505 & $56\pm1$ & $320\pm60$ & $7970\pm330$ & $7369\pm108$\\
			184917633 & $56\pm1$ & $270\pm80$ & $7780\pm400$ & $7443\pm122$\\
			
    \end{tabular}
    }
	\label{tb:indices}
\end{table}

\section{The grids of models}\label{sec:grid}

As $\delta$ Sct are usually moderate to fast rotators ($v\sin i$ > 100 \kms), the models have to take into account the stellar structure deformation that occurs at these speeds. This centrifugal flattening reduces the value of the mean density, directly related to one of the seismic indices that we are using here, the large separation. For this reason, we compute our models with the \mesa\ code, version 15140, \citep{Paxton2019}, and the related oscillations with the \filou\ code \citep{SG2008}, taking rotation into account up to second order in the perturbative theory for the adiabatic oscillation computation (including near-degeneracy effects and stellar structure deformation). 

Following Paper~I, we build two grids of representative models to characterise \dss\ during their stay on the Pre-Main Sequence (PMS) and the MS, one for Trumpler 10 and another for Praesepe. In Table~\ref{tb:grid} we introduce their main parameters. We used the default nuclear reactions network, basic.net, provided by \mesa. At the zero-age main sequence (ZAMS), X being constant, when Z increases, Y decreases by the same amount, so that X + Y + Z = 1. After testing exponential overshooting values $f_{0} = 0.002$ and $f = 0.022$, we find no significant impact and thus use no overshooting. We include internal differential rotation. We computed models initiating rotation at the ZAMS. The values for the initial angular velocity to critical velocity ratio are between $0.1\Omega_{c}$ and $0.5\Omega_{c}$, avoiding higher values that may lie beyond the limits of the perturbative theory. 
We compute $p$ modes between $n=1$ and the cut-off frequency, the limiting frequency up to which acoustic modes can propagate without damping. We use the modes in the low-order regime, between $n=2$ and $n=8$, and with degrees between $l=0$ and $l=3$, to calculate the low-order large separation, as proposed by \citet{Suarez2014}.

\begin{table}
	\centering
	\caption{Parameters of the stellar model grids built with the \mesa\ code. From top to bottom: mass, metallicity (both for Trumpler 10 and Praesepe), the initial angular velocity to critical velocity ratio and the mixing-length parameter.}
    \renewcommand{\arraystretch}{1.3}
	\addtolength{\tabcolsep}{8 pt}
	\resizebox{7cm}{!}{
    \begin{tabular}{ccc}
			\hline
			Parameter & Range & Step\\
			\hline
			$M ({\rm M}_{\odot})$ & [1.60, 2.50] & 0.01 ${\rm M}_\odot$\\
			Z (Trumpler 10) & [0.016, 0.020] & 0.002\\
			Z (Praesepe) & [0.028, 0.032] & 0.002\\
			$\Omega / \Omega_{c}$ & [0.1, 0.5] & 0.1\\
			$\alpha$ & 2.0 & Fixed\\
			
    \end{tabular}
    }
	\label{tb:grid}
\end{table}

\section{The method for estimating ages}\label{sec:mode_identification}

We estimate the seismic age of each cluster following these steps:

\begin{enumerate}
\item
For each star, we constrain the models using its estimated values of \Dnulow\ and $\tilde{T}_\mathrm{eff}$, taking into account their corresponding uncertainties.
\item
We compute the ratios of the observed frequencies, in order to select the frequencies with ratios that match those of the models (Table \ref{tb:relations}). 
\item
Once we find a range of radial modes matching the observed frequencies, we use them with the models selected at step 1, to better constrain the stellar ages.
\item
After applying steps 1-3 to all stars in the same group, we plot the age distribution weighted histograms of all the constrained models. 
\item
We compute the best weighted probability density function (WPDF) over the histogram. For that, we assume a normal distribution, and we use, as initial guesses, the maximum likelihood estimation: the weighted mean of all the ages of the constrained models, and its corresponding weighted standard deviation.  We finally take, for the age of the cluster and its corresponding standard deviation, the parameters of the fitted WPDF in a $\chi^{2}$ minimisation process.  
\end{enumerate}

We compute the weight of each constrained model, $p$ given in Eq.~(\ref{eq4}), taking into account the following assumptions:
\begin{enumerate}
    \item Each evolutionary track computed with \mesa\ is oversampled at low ages. To compensate for this effect, our WPDF is proportional to the time step, $\Delta t$, divided by the total time, $t$, of the model in its evolutionary track. 
    \item We also evaluate the probability of the models corresponding to each star in the sample. Models corresponding to stars with better measured \Dnulow\ and  $\tilde{T}_\mathrm{eff}$ contribute with a greater probability in the estimated age of the cluster. Then the WPDF is inversely proportional to the relative uncertainties of $\frac{e_{\Delta\nu_{\rm low}}}{\Delta\nu_{\rm low}}$ and  $\frac{e_{\Tilde{T}_\mathrm{eff}}}{\Tilde{T}_\mathrm{eff}}$ of each star.
    \item Regarding the probability that a model is the age of the whole cluster, we assign it a weight proportional to the number of stars that are this same age, $n_{\rm stars}$ divided by the total number of stars in the sample, ${N_{\rm stars}}$.
    \item The number of constrained models with the same age, $n$, divided by the total number of models in the grid, $N$.
\end{enumerate}

Combining everything, we obtain
\begin{equation}
    p = \frac{\Delta t}{t} \frac{\Delta\nu_{\rm low}}{e_{\Delta\nu_{\rm low}}} \frac{\Tilde{T}_\mathrm{eff}}{e_{\Tilde{T}_\mathrm{eff}}} \frac{n_{\rm stars}}{N_{\rm stars}} \frac{n}{N}.\label{eq4}
\end{equation}

The formula for $\chi^{2}$ (Eq. ~\ref{eq5}) has been applied over the densities of the histogram, his(age), in order to obtain the best possible fit to a normal distribution, norm(age):
\begin{equation}
    \chi^{2} = \sum_{\rm bins}{\frac{\rm ( his(age)-norm(age) )^2}{\rm norm(age)^{2}}}.\label{eq5}
\end{equation}

\begin{table}
	\centering
	\caption{The fundamental mode to radial overtone ratios in our \mesa/\filou\ grids of models, with their corresponding standard deviations.}
    \renewcommand{\arraystretch}{1.3}
	\addtolength{\tabcolsep}{8 pt}
	\resizebox{6 cm}{!}{
    \begin{tabular}{ccc}
			\hline
			Relationship & Value with 1$\sigma$ uncertainty\\
			\hline
			$f_{1}/f_{2}$ & $0.77\pm0.01$\\
			$f_{1}/f_{3}$ & $0.63\pm0.02$\\
			$f_{1}/f_{4}$ & $0.53\pm0.02$\\
			$f_{1}/f_{5}$ & $0.45\pm0.02$\\
			$f_{1}/f_{6}$ & $0.40\pm0.02$\\
			$f_{1}/f_{7}$ & $0.35\pm0.01$\\
			$f_{1}/f_{8}$ & $0.31\pm0.01$\\
    \end{tabular}
}
	\label{tb:relations}
\end{table}

\begin{figure*}   
    \centering
    \includegraphics[width=0.8\textwidth]{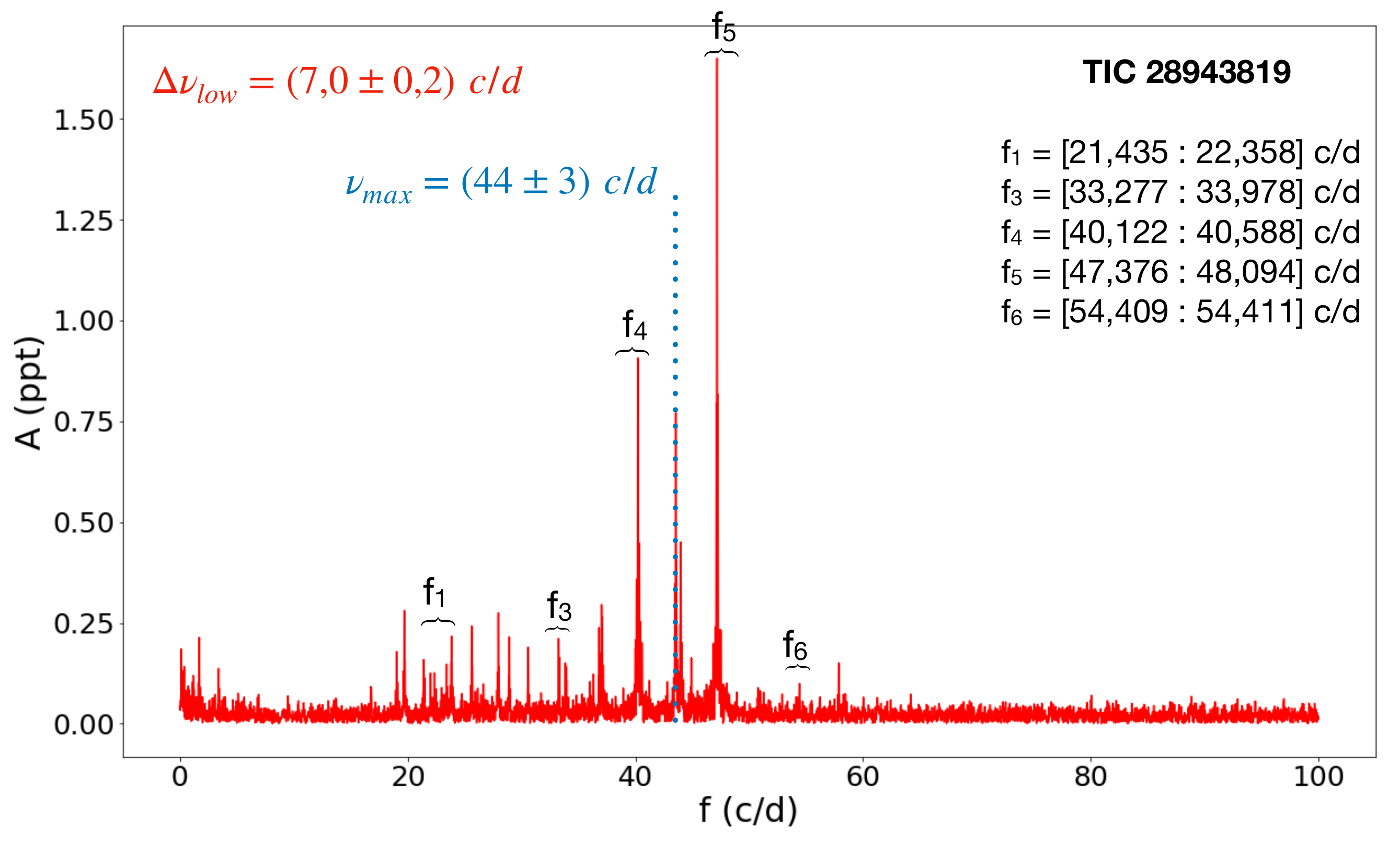}
    \caption{Ranges for the possible radial overtones in the frequency spectrum of TIC\,28943819.}
    \label{fig:TIC28943819_radial_overtones}
\end{figure*}

Fig.~\ref{fig:TIC28943819_radial_overtones} show the positions and the ranges of the possible radial overtones in the frequency spectrum of TIC\,28943819 (see Appendix~\ref{appendixC} for the rest of our sample). These ranges are too wide in some cases because we have sampled the whole grid for identification. Then, the inclusion of the fundamental mode has a minimal impact on the constraints we derive on the models, but it helps confirm what we obtain from the other seismic parameters. The identification has failed estimating the ranges for the possible radial overtones only in the cases with fewer than 30 extracted significant frequencies: TIC\,175194881 in Praesepe, and TIC\,30307085 and TIC\,271061334 in Trumpler 10. That's why in Appendix~\ref{appendixC} there are seven figures, instead of 10, the total number of stars in our sample, plus TIC\,28943819.

\section{Results and discussion}\label{sec:results}

\subsection{Trumpler 10}
\begin{figure*}   
    \centering
    \includegraphics[width=0.9\textwidth]{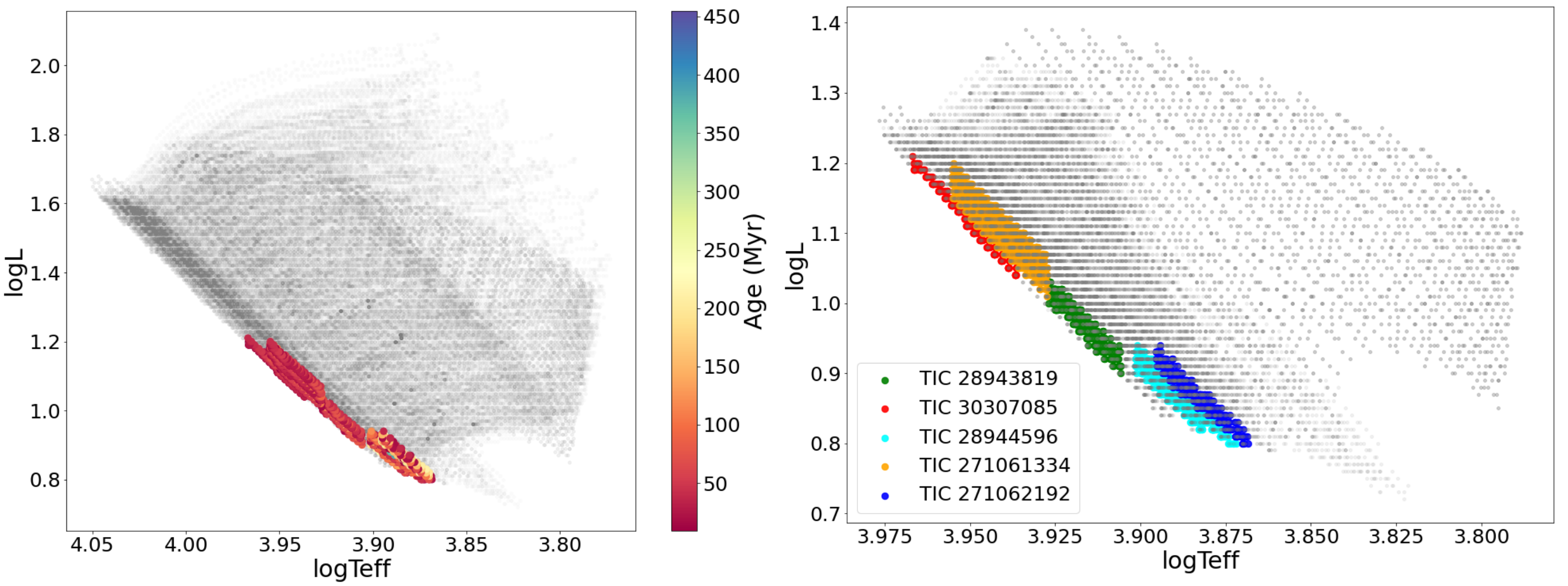}
    \caption{Left: HRDs of the evolutionary tracks of our grid of representative models for the sample of \dss\ in Trumpler 10, the ages of the seismically constrained models have been colour coded. Right: Zoom between 1.60 $M_{\odot}$ and 2.00 $M_{\odot}$, distinguishing between models for each of the stars in the sample by different colours.}
    \label{fig:Trumpler10_HR}
\end{figure*}

The HRD of Fig.~\ref{fig:Trumpler10_HR} (left panel) shows the ages of the seismically constrained models of our sample of \dss\ in Trumpler 10. We can see that they are very close to the ZAMS. Focussing on models between 1.60 and 2.00 ${M_{\odot}}$~(Fig.~\ref{fig:Trumpler10_HR}, right panel), the constrained models show that TIC\,28944596 and TIC\,271062192, the least massive stars of the sample, seem to be older than the rest of the sample. Maybe both stars are actually older, or maybe these apparent older ages have to do with a gravity-darkening effect or a possible binary companion. The observed larger radii, the lower densities and higher luminosities (presented in Table~\ref{tb:parametersTr10}) could explain the three hypotheses. 

\begin{figure*}   
    \centering
    \includegraphics[width=0.9\textwidth]{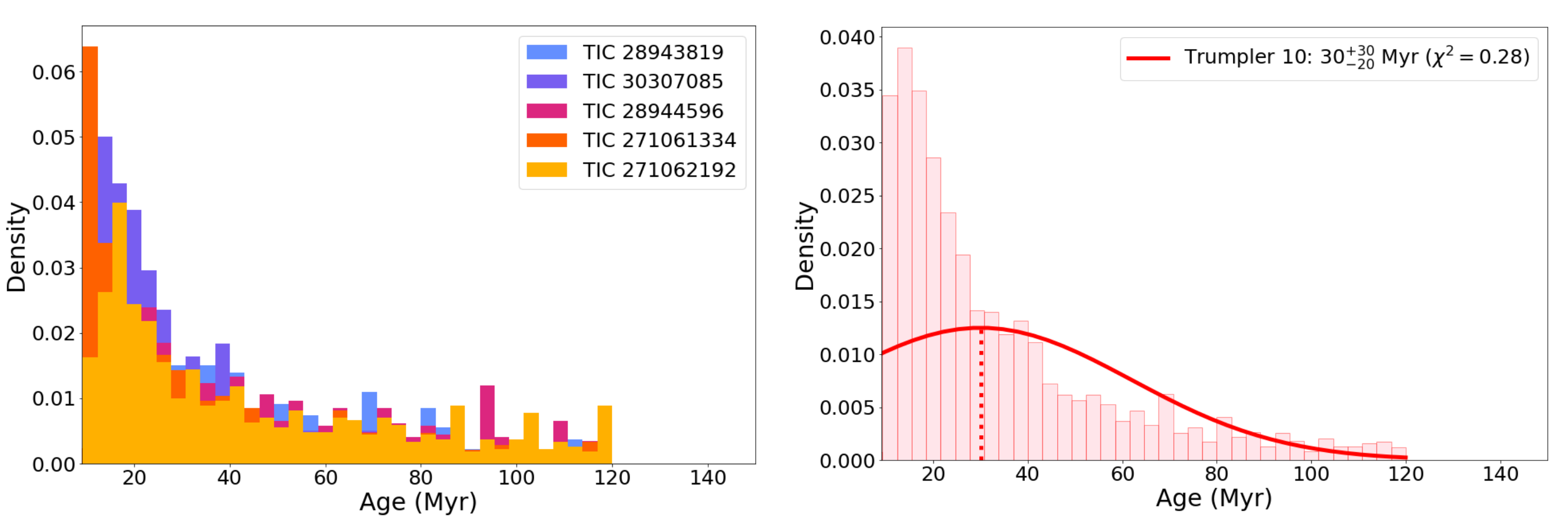}
    \caption{Age weighted histograms of our sample of stars in Trumpler 10. Left: Weighted histogram of each member of the sample. Right: Weighted histogram of the whole sample. The red solid line shows the computed PDF with a mean age of $30_{-20}^{+30}$ \myr.}
    \label{fig:Trumpler10_age_distribution}
\end{figure*}

To determine the mean age of the group, we first compute the age weighted histograms corresponding to every star of the sample, from the seismically constrained models (Fig.~\ref{fig:Trumpler10_age_distribution}, left panel). We then estimate the mean age of the whole group, as a single population, by calculating the best possible distribution on the histograms, using a normal WPDF, as explained in Sec.~\ref{sec:mode_identification} (Fig.~\ref{fig:Trumpler10_age_distribution}). The result is a mean age of around $30_{-20}^{+30}$ \myr, very close to the ZAMS. This is a younger age estimate than those referenced in Sect.~\ref{sec:ages}, compatible with estimates of \citet{Kharchenko2013,Netopil2016}, of around 34 \myr\ and 40 \myr, respectively. Uncertainties probably emerge because seismic parameters, including the large separation, evolve rapidly for stars on the PMS.

Recent theoretical works show that the PMS is a complex phase. For example, \citet{Kunitomo2017} claim that the spread in luminosity during the PMS can be explained through different efficiencies at which the accreted material is converted into internal energy for each star. For these very young clusters, it seems that we need other parameters, in addition to the seismic ones we use, to date the stars with greater accuracy. This is confirmed by \citet{Steindl2022}, according to whom 
different PMS accretion scenarios cause differences in the pulsation modes, thus leaving an imprint on the frequency content of a $\delta$ Sct star. Seismology of PMS stars has a lot to say about their interior structure. 

Compared to the works of \citet{Murphy2021,Steindl2022}, our uncertainties are one order of magnitude larger with the stars of Trumpler 10. \citet{Murphy2021} compute {\Dnulow\ $= 6.83\pm0.01\ d^{-1}$} for the PMS star HD\,139614, a value one order of magnitude more precise than our measured large separations. By scanning a variety of models for mode identification, we sample the entire relevant parameter space, which makes uncertainties larger but more realistic. We limit our mode identification precision by avoiding over-reliance on our models.

\begin{table*}
	\centering
	\caption{Constrained parameters of the models corresponding to our selected targets in Trumpler 10, with their corresponding standard deviations.}
    \renewcommand{\arraystretch}{2}
	\addtolength{\tabcolsep}{2 pt}
	\resizebox{13cm}{!}{
    \begin{tabular}{ccccccccccc}
			\hline
			TIC & $M (M_{\odot})$ & $R (R_{\odot})$ & $\bar \rho (\bar \rho_{\odot})$ & \logg & $\tilde{T}^\mathrm{eff}(K)$ & $log(L/L_{\odot})$ & $v_{rot}$(\kms) & Age (\myr) \\
			\hline
			28943819 & $1.72\pm0.04$ & $1.51\pm0.02$ & $0.50\pm0.02$ & $4.31\pm0.01$ & $8250\pm120$ & $0.97\pm0.03$  & $70_{-40}^{+40}$ & $30_{-20}^{+30}$  \\
			30307085 & $1.87\pm0.05$ & $1.53\pm0.02$ & $0.52\pm0.01$ & $4.34\pm0.01$ & $8920\pm180$ & $1.12\pm0.04$  & $60_{-10}^{+20}$ & $20_{-10}^{+20}$  \\
			28944596 & $1.64\pm0.03$ & $1.51\pm0.02$ & $0.48\pm0.02$ & $4.29\pm0.01$ & $7780\pm120$ & $0.87\pm0.03$  & $80_{-40}^{+40}$ & $40_{-30}^{+40}$  \\
			271061334 & $1.87\pm0.05$ & $1.58\pm0.03$ & $0.48\pm0.02$ & $4.31\pm0.01$ & $8740\pm160$ & $1.11\pm0.04$  & $80_{-50}^{+50}$ & $20_{-15}^{+30}$  \\
			271062192 & $1.64\pm0.03$ & $1.55\pm0.02$ & $0.44\pm0.02$ & $4.27\pm0.01$ & $7670\pm120$ & $0.87\pm0.03$ & $110_{-50}^{+50}$ & $40_{-30}^{+40}$  \\
			
    \end{tabular}
    }
	\label{tb:Trumpler10_models}
\end{table*}

\begin{table*}
	\centering
	\caption{Same as Table~\ref{tb:Trumpler10_models} for the Praesepe sample.}
    \renewcommand{\arraystretch}{2}
	\addtolength{\tabcolsep}{2 pt}
	\resizebox{13cm}{!}{
    \begin{tabular}{ccccccccccc}
			\hline
			TIC & $M (M_{\odot})$ & $R (R_{\odot})$ & $\bar \rho (\bar \rho_{\odot})$ & \logg & $\tilde{T}^\mathrm{eff}(K)$ & $log(L/L_{\odot})$ & $v_{rot}$(\kms) & Age (Myr) \\
			\hline
			175194881 & $1.83\pm0.03$ & $1.92\pm0.02$ & $0.26\pm0.01$ & $4.14\pm0.01$ & $7680\pm100$ & $1.06\pm0.03$  & $130\pm60$ & $470\pm120$  \\
			175264376 & $1.73\pm0.05$ & $2.03\pm0.05$ & $0.21\pm0.01$ & $4.06\pm0.02$ & $7190\pm160$ & $0.99\pm0.05$  & $110\pm50$ & $880\pm130$  \\
			175265807 & $1.85\pm0.04$ & $1.94\pm0.03$ & $0.25\pm0.01$ & $4.13\pm0.01$ & $7730\pm130$ & $1.08\pm0.04$  & $140\pm60$ & $480\pm130$  \\
			175291778 & $1.72\pm0.06$ & $2.02\pm0.06$ & $0.21\pm0.02$ & $4.06\pm0.02$ & $7150\pm180$ & $0.98\pm0.06$  & $120\pm50$ & $890\pm150$  \\
			184914505 & $1.82\pm0.05$ & $1.96\pm0.03$ & $0.24\pm0.01$ & $4.11\pm0.01$ & $7590\pm160$ & $1.05\pm0.05$ & $130\pm60$ & $580\pm130$  \\
			184917633 & $1.77\pm0.06$ & $1.93\pm0.03$ & $0.25\pm0.01$ & $4.11\pm0.01$ & $7430\pm200$ & $1.01\pm0.06$ & $140\pm50$ & $600\pm160$  \\
			
    \end{tabular}
    }
	\label{tb:Praesepe_models}
\end{table*}

\subsection{Praesepe}
\begin{figure*}   
    \centering
    \includegraphics[width=0.9\textwidth]{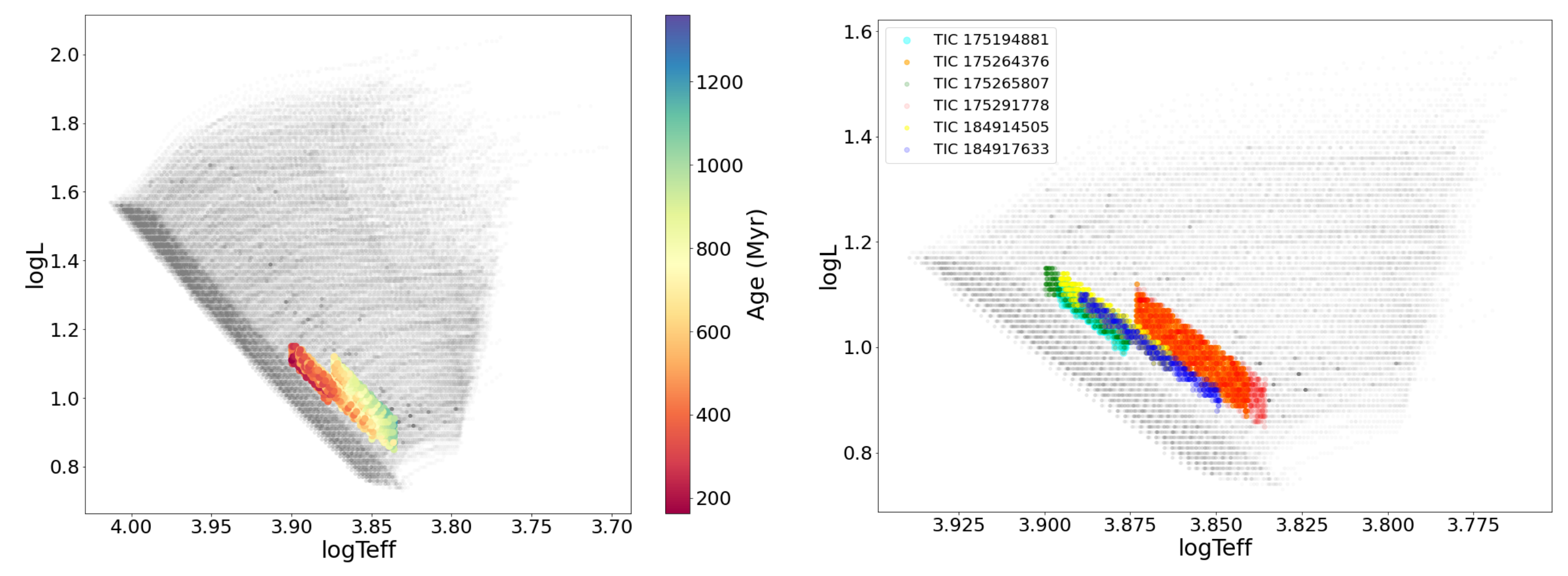}
    \caption{The same as Fig.~\ref{fig:Trumpler10_HR} for Praesepe.}
    \label{fig:Praesepe_HR}
\end{figure*}

\begin{figure*}   
    \centering
    \includegraphics[width=0.9\textwidth]{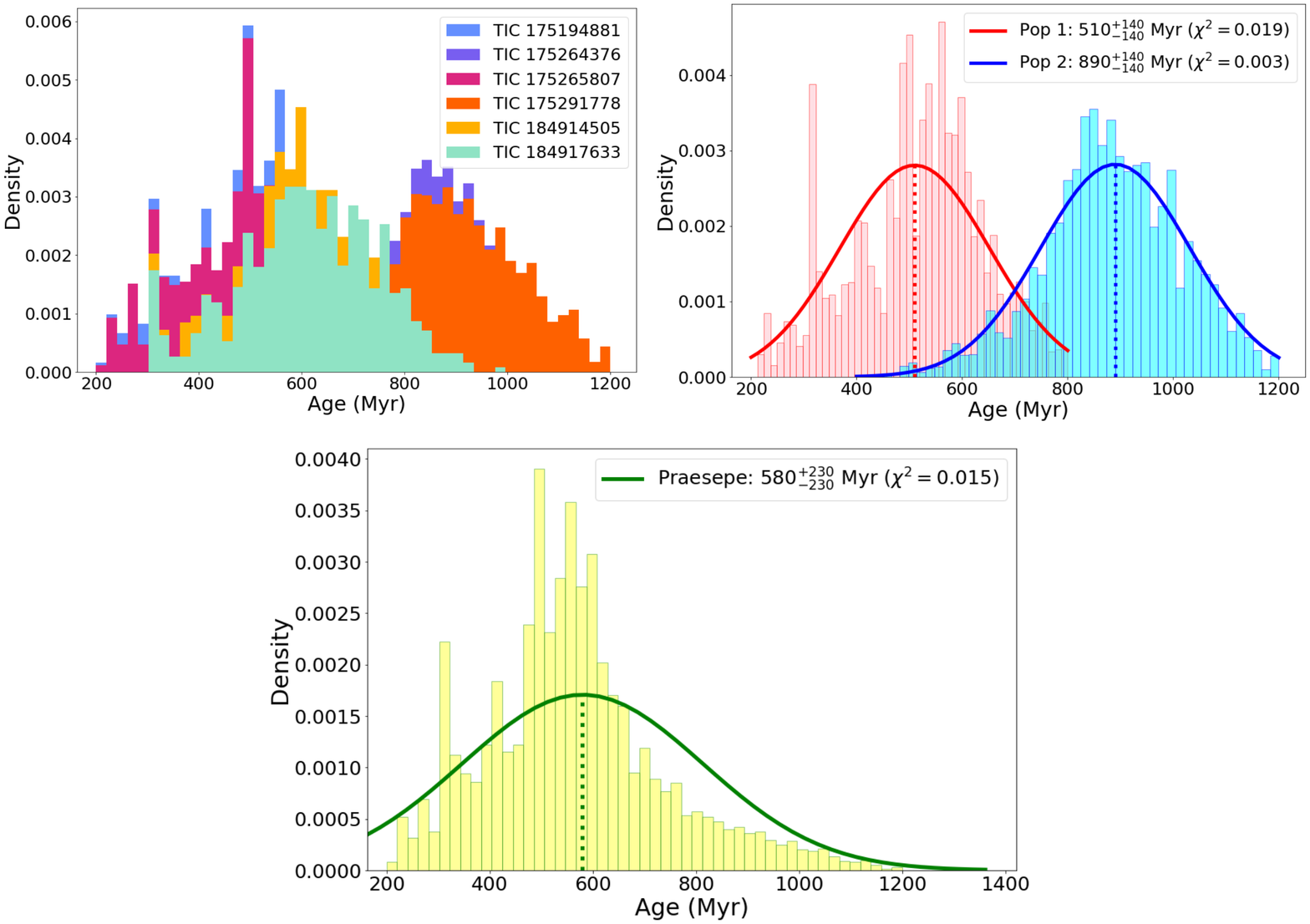}
    \caption{Age weighted histograms of the seismically constrained models of our sample of stars in Praesepe 10. Top left: Weighted histograms for each member of the sample. Top right: Weighted histograms for the apparent two \dss\ populations: Pop 1 groups the stars TIC\,175194881, TIC\,175265807, TIC\,184914505 and TIC\,184917633, and Pop 2 groups the stars TIC\,175264376 and TIC\,175291778. The red solid line is the computed PDF of Pop 1, with a mean age of $510\pm140$ \myr. The blue solid line is the computed PDF of Pop 2, with a mean age of $890\pm140$ \myr. Bottom: Weighted histogram of the whole sample, considered as one single population. The green solid line is the computed PDF, with a mean age of $580\pm230$ \myr.}
    \label{fig:Praesepe_KDE_distribution}
\end{figure*}

Fig.~\ref{fig:Praesepe_HR} shows the HRD of the seismically constrained models for the \dss\ group in Praesepe. Two stars, TIC\,175264376 and TIC\,175291778, clearly appear to be older than the rest of the sample (right panel). This is more evident in Fig.~\ref{fig:Praesepe_KDE_distribution} (top left panel), where we plot the age weighted histogram for every star of the sample. If we consider the sample as a single population, then the computed WPDF (bottom panel) give us a mean age of $580\pm230$~\myr, younger if we compare it to the references cited in Sect.~\ref{sec:ages}, but in good agreement with them. It is significant that our estimate is very near of the age used by \citet{Fossati2008}, $590_{-120}^{+150}$ \myr, where they calculated the metallicity of the cluster from an abundance analysis of A- and F-type stars. Five of them have been used in the present work. 

The discrepant large separations and densities of TIC\,175264376 and TIC\,175291778 can be explained through rotation, a different age, binary systems or less trustful estimated seismic parameters. First, a rapid rotation may modify the value of the large separation, although not the scaling relation between the large separation and the mean density \citep{GH2015,Mirouh2019}.  Their relatively higher TIC radii and lower TIC densities (Table~\ref{tb:parametersM44}), are in line with the lower low-order large separations and frequencies at maximum power we estimate for both stars (Table~\ref{tb:indices}). The high value of the projected rotation velocity of TIC\,175264376 (200 \kms) is very significant in this sense. Our 1D models cannot be reliably applied to such high rotational velocities. We then need 2D models to characterise rapidly-rotating stars. In a more evolved cluster as Praesepe, rotation and other internal mixing phenomena can affect the stars differently over time. 

Second, these bigger and lower-density stars with masses similar to the others, can also simply be more evolved. In Table~\ref{tb:parametersM44} we can see that these stars, plus TIC\,175194881, may be late A or earlier F stars, while the other three may be middle A stars, according to \citet{Fossati2008}. These could point to two different populations of stars. 

Third, the outlier stars may be in binary systems. Their luminosities would be brighter than models of single stars would suggest, making them appear more evolved. 

Then, we revisit our one-population assumption, as there could be two populations, in which the four younger stars would be 'Pop 1' and TIC\,175264376 and TIC\,175291778 would be 'Pop 2' (Fig.~\ref{fig:Praesepe_KDE_distribution}, top right panel). Once the WPDF of both populations are computed, we obtain a mean age of $510\pm140$ \myr\ for `Pop 1', and a mean age of $890\pm140$ \myr\ for `Pop 2'. These histograms are quite different from the histogram with a single population, which brings us to a fourth explanation for this apparent bimodality. Compared to `Pop 1' stars, those in `Pop 2' contribute less weight to the WPDF of the totality of the constrained models, due to their larger uncertainties in the measurements of the low-large separation and the seismic effective temperature. Then, assuming one single population, the age of the cluster is closer to the age of `Pop 1' than the age of `Pop 2'. 

As we can see in Tables~\ref{tb:Trumpler10_models} and~\ref{tb:Praesepe_models}, the models are not well constrained in terms of rotational velocity. Estimating the rotation rate of some stars of the group would yield further constraints on the models, especially given the dependence of the seismic parameters such as the large separation or the frequency ratios on rotation \citep{Suarez2006}. 

\citet{Murphy2022} use a grid of non-rotating stars to model the three slowest rotators from the Pleiades sample, in order to verify their mode identification. However, they do not model the other two stars, rapid rotators V1228 Tau (vsini = 200 \kms) and V650 Tau (vsini = 230 \kms), for which the échelle diagrams are more ambiguous. They also conclude that rotating models are required for a more accurate inference of the asteroseismic parameters, including the mode identification. A high rotation mixes the modes in such a way that it is not easy to identify them within such a complex spectrum. In this analysis, and that of Paper~I, we have included rotating models and defined a strategy that will serve as a stepping stone towards a complete mode identification in rapid rotators. Then, to advance this strategy, we need a method to help us confidently interpret the rotation frequency in the dense frequency spectrum of \dss. It is also crucial to obtain a more reliable mode identification, through longer exposure observations that allow a higher resolution in the frequency spectrum. We hope that future missions, such as the ESA projects PLATO\footnote{https://sci.esa.int/web/plato} \citep[PLAnetary Transits and Oscillations of stars,][]{Rauer2014} and HAYDN\footnote{http://www.asterochronometry.eu/haydn/} \citep[High-precision AsteroseismologY in DeNse stellar fields,][]{Miglio2021}, will provide higher resolution photometry of stars belonging to clusters, leading to accurate age estimates.

\section{Conclusions}\label{sec:conclusions}

The use of asteroseismology to date young open clusters provides promising results, despite the limitations of statistical techniques applied to samples with such small numbers of stars. We have tested a seismic method with three open clusters of different ages. With $\alpha$ Per we obtained the first results in \citet{PamosOrtega}. In this work we extend the research to Trumpler 10 and Praesepe, and we estimate the ranges with the possible values of some radial overtones and the fundamental mode. 

Regarding Trumpler 10, we find five $\delta$ Sct star candidates never before classified as such, with a mean age of $30_{-20}^{+30}$ \myr. The uncertainty is large due to how close they are to the PMS, where stars evolve rapidly. Other parameters are needed to better constrain the models near the ZAMS, and to more accurately date stars.  

Regarding Praesepe, we find a new possible $\delta$ Scuti star, TIC\,184917633, that we add to our sample of five previously known \dss. We estimate the mean age of this star group to $580\pm230$ \myr, in good agreement with the literature. Two of the six stars in the sample seem to be older than the rest. The different values in their parameters, especially the spectral type, support the thesis of two stellar populations: one with a mean age of $510\pm140$ \myr\ and another with a mean age of $890\pm140$ \myr. This apparent bimodality in the age distribution could also be due to the effects of gravity-darkening in rapidly-rotating stars. The lower values in the low-order large separation and the frequency at maximum power, in addition to the measured large projected rotation velocity of both stars, support this idea. The 1D models that we have used in this work are not the most suitable to stars with such high rotation rates. Two-dimensional models are needed in order to take more into account the deformation that occurs in them, and that greatly impact the seismic parameters, such as the large separation and a reliable determination of the rotation frequency. Finally, we cannot rule out an unreliable estimate of the seismic parameters of the outlier stars. A greater weight of the other four stars in the WPDF corrects for this apparent bimodality.

\begin{acknowledgements}
We appreciate the comments and questions from the anonymous referee, because they have contributed to improving the paper. 
DPO and AGH acknowledge funding support from ‘FEDER/junta de Andalucía-Consejería de Economía y Conocimiento’ under project E-FQM-041-UGR18 by Universidad de Granada. 
JCS, GMM and SBF acknowledge funding support from the Spanish State Research Agency (AEI)  project PID2019-107061GB-064.  
This paper includes data collected with the TESS mission, obtained from the TASC data archive. Funding for the TESS mission is provided by the NASA Explorer Program. STScI is operated by the Association of Universities for Research in Astronomy, Inc., under NASA contract NAS 5–26555.

\end{acknowledgements}

\bibliographystyle{aa}
\bibliography{bibl}    
\clearpage

\begin{appendices}
\setcounter{table}{0} 
\renewcommand{\thetable}{A.\arabic{table}}

\section{The ten modes with highest extracted amplitudes and associated frequencies for each star in our $\delta$ Sct star sample.}
\label{appendixA}

	\renewcommand{\arraystretch}{1.3}
    \resizebox{7cm}{!}{
    \begin{tabular}{ccc|ccc}
			\hline
			\multicolumn{3}{c|}{Trumpler 10} & \multicolumn{3}{c}{Praesepe}\\
			\hline
			TIC & $f (d^{-1})$ & A (ppt) & 
			TIC & $f (d^{-1})$ & A (ppt) \\
			\hline
			\multirow{10}{*}{TIC\,28943819} & 47.185 & 1.667
			                               &        
			                               & 31.566 & 0.645\\
			                               & 40.230 & 0.885
			                               & 
			                               & 30.242 & 0.314\\
			                               & 43.554 & 0.787
			                               &               
			                               & 29.507 & 0.426\\
			                               & 44.008 & 0.463
			                               &               
			                               & 31.896 & 0.351\\
			                               & 40.122 & 0.342
			                               &               
			                               & 27.811 & 0.168\\
			                               & 40.359 & 0.335
			                               & TIC\,175194881
			                               & 26.556 & 0.117\\
			                               & 37.046 & 0.318
			                               &               
			                               & 36.742 & 0.079\\
			                               & 19.721 & 0.282
			                               &               
			                               & 21.457 & 0.081\\
			                               & 27.977 & 0.277
			                               &               
			                               & 23.489 & 0.051\\
			                               & 36.830 & 0.251
			                               &               
			                               & 38.623 & 0.047\\
			\hline
			\multirow{10}{*}{TIC\,30307085} & 66.752 & 2.630
			                               &               
			                               & 14.845 & 1.471\\
			                               & 62.280 & 2.009
			                               &               
			                               & 12.021 & 0.914\\
			                               & 55.489 & 1.809
			                               &               
			                               & 19.260 & 1.055\\
			                               & 53.359 & 0.810
			                               &               
			                               & 22.514 & 0.971\\
			                               & 59.387 & 0.703
			                               &               
			                               & 27.996 & 0.919\\
			                               & 60.692 & 0.548
			                               & TIC\,175264376
			                               & 19.167 & 0.610\\
			                               & 57.244 & 0.448
			                               &               
			                               & 15.575 & 0.511\\
			                               & 4.427 & 0.359
			                               &              
			                               & 13.359 & 0.394\\
			                               & 48.369 & 0.344
			                               &               
			                               & 10.180 & 0.339\\
			                               & 64.533 & 0.292
			                               &               
			                               & 16.501 & 0.292\\
			\hline
			\multirow{10}{*}{TIC\,28944596} & 25.341 & 1.674
			                               &               
			                               & 35.980 & 1.629\\
			                               & 34.691 & 1.588
			                               &               
			                               & 31.020 & 1.618\\
			                               & 24.494 & 1.358
			                               &               
			                               & 28.457 & 1.200\\
			                               & 33.649 & 1.089
			                               &               
			                               & 24.659 & 0.975\\
			                               & 16.494 & 0.815
			                               &               
			                               & 29.333 & 0.949\\
			                               & 39.021 & 0.682
			                               & TIC\,175265807
			                               & 33.417 & 0.829\\
			                               & 17.145 & 0.535
			                               &               
			                               & 33.544 & 0.812\\
			                               & 35.790 & 0.513
			                               &               
			                               & 26.086 & 0.708\\
			                               & 22.821 & 0.398
			                               &               
			                               & 33.794 & 0.300\\
			                               & 32.201 & 0.344
			                               &               
			                               & 28.901 & 0.262\\
			\hline
			\multirow{10}{*}{TIC\,271061334} & 60.529 & 1.163
			                                &               
			                                & 23.855 & 2.336\\
			                                & 48.627 & 0.684
			                                &               
			                                & 22.667 & 1.093\\
			                                & 58.877 & 0.474
			                                &               
			                                & 7.908 & 0.790\\
			                                & 50.485 & 0.455
			                                &               
			                                & 20.512 & 0.702\\
			                                & 53.414 & 0.201
			                                &               
			                                & 7.332 & 0.640\\
			                                & 55.330 & 0.192
			                                & TIC\,175291778                  & 24.209 & 0.631\\
			                                & 53.583 & 0.168
			                                &               
			                                & 6.827 & 0.612\\
			                                & 59.792 & 0.147
			                                &               
			                                & 12.141 & 0.609\\
			                                & 60.696 & 0.134
			                                &               
			                                & 18.050 & 0.546\\
			                                & 0.212 & 0.157
			                                &              
			                                & 17.146 & 0.519\\
			\hline
			\multirow{10}{*}{TIC\,271062192} & 21.680 & 1.397
			                                &               
			                                & 29.667 & 1.849\\
			                                & 18.611 & 0.703
			                                &               
			                                & 31.899 & 1.543\\
			                                & 29.522 & 0.367
			                                &               
			                                & 29.465 & 1.138\\
			                                & 43.681 & 0.351
			                                &               
			                                & 26.145 & 1.168\\
			                                & 42.380 & 0.238
			                                &               
			                                & 31.565 & 1.059\\
			                                & 33.928 & 0.224
			                                & TIC\,184914505
			                                & 29.257 & 0.869\\
			                                & 3.527 & 0.209
			                                &              
			                                & 25.231 & 0.782\\
			                                & 7.028 & 0.178
			                                &              
			                                & 12.482 & 0.591\\
			                                & 59.146 & 0.157
			                                &               
			                                & 12.233 & 0.417\\
			                                & 5.777 & 0.154
			                                &              
			                                & 27.690 & 0.323\\
		    \hline
		    \multirow{10}{*}{            }  &        & 
		                                    &               
		                                    & 13.309 & 1.359\\
		                                    &        & 
		                                    &       
		                                    & 30.241 & 1.019\\
		                                    &        &
		                                    &
		                                    & 27.305 & 0.613\\
		                                    &        &      
		                                    &
		                                    & 16.610 & 0.581\\
		                                    &        &    
		                                    &
		                                    & 26.031 & 0.531\\
		                                    &        &
		                                    & TIC\,184917633
		                                    & 32.840 & 0.526\\
		                                    &        &
		                                    &
		                                    & 26.241 & 0.518\\
		                                    &        &
		                                    &
		                                    & 22.494 & 0.435\\
		                                    &        &
		                                    &
		                                    & 13.192 & 0.402\\
		                                    &        &
		                                    &
		                                    & 28.240 & 0.409\\
    \hline
    \end{tabular}
    }
    \captionof{table}{Ten highest-amplitude modes and their frequencies, for each star in our samples of Trumpler 10 and Praesepe. The tables, corresponding to each star in our sample, with all the extracted frequencies is available on-line at CDS}
	\label{tb:frequencies}

\clearpage

\section{The estimated large separations of our $\delta$ Sct star sample}
\label{appendixB}

\setcounter{figure}{0} 
\renewcommand{\thefigure}{B.\arabic{figure}}

\includegraphics[width=0.7\textwidth,]{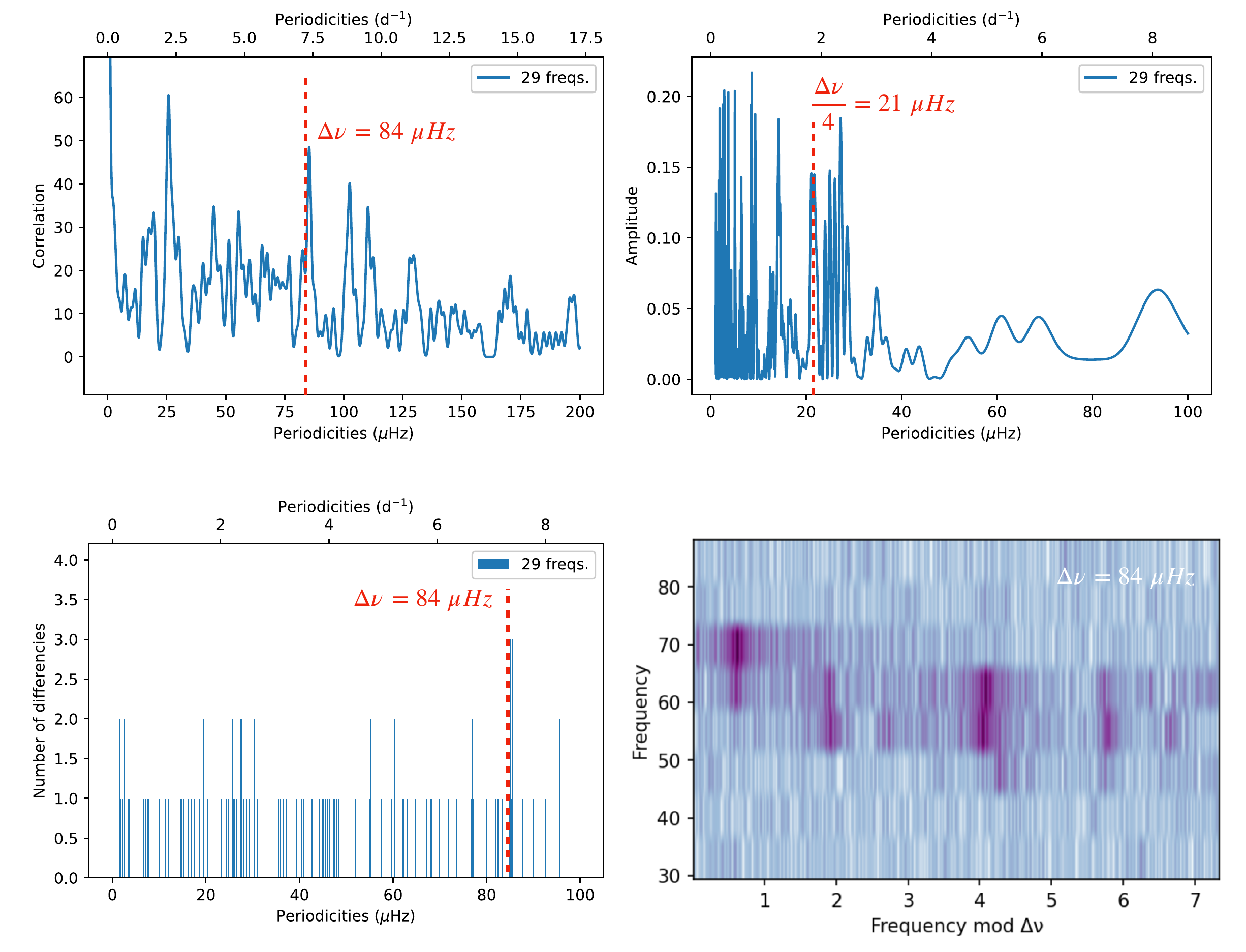}
\captionof{figure}{The same as  Fig.~\ref{fig:TIC28943819_large_separation} for TIC\,30307085.}
\captionsetup{}
\label{fig:30307085_large_separation} 
\hfill \break
\hfill \break

\includegraphics[width=0.7\textwidth,]{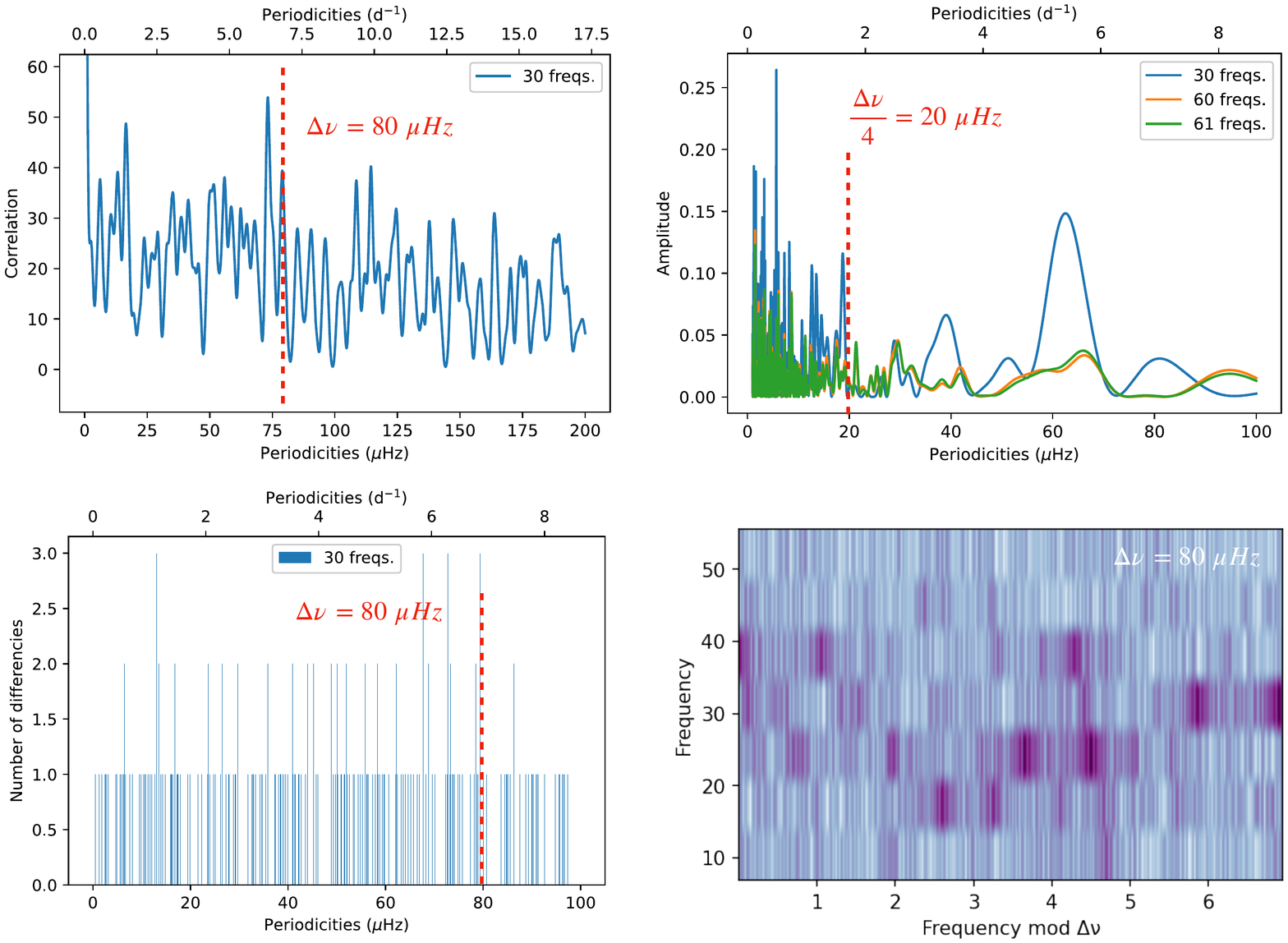}
\captionof{figure}{The same as  Fig.~\ref{fig:TIC28943819_large_separation} for TIC\,28944596.}
\label{fig:28944596_large_separation}

\clearpage

\includegraphics[width=0.7\textwidth]{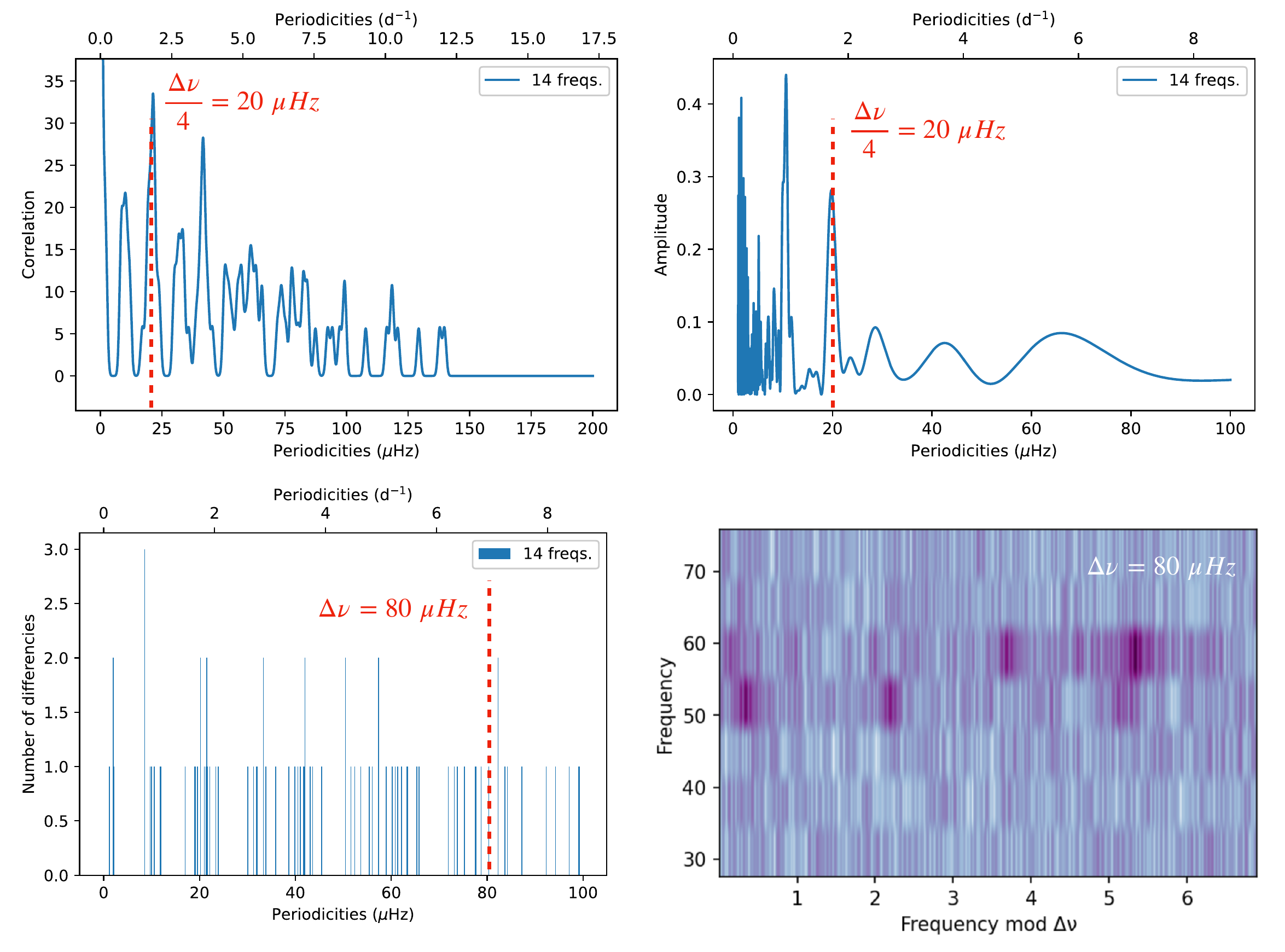}
\caption{The same as  Fig.~\ref{fig:TIC28943819_large_separation} for TIC\,271061334.}
\label{fig:271061334_large_separation}
\hfill \break
\hfill \break

\includegraphics[width=0.7\textwidth]{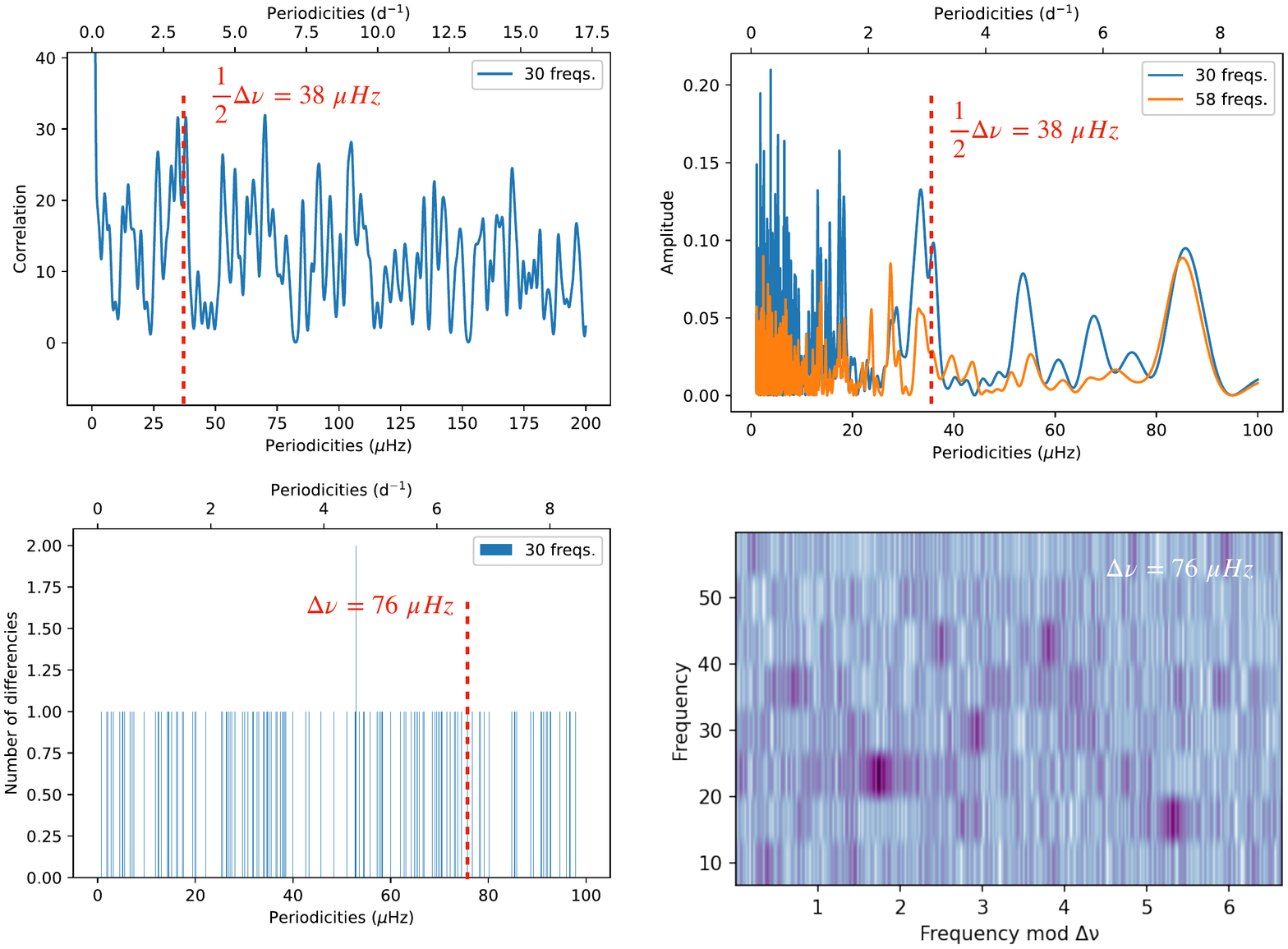}
\caption{The same as  Fig.~\ref{fig:TIC28943819_large_separation} for TIC\,271062192.}
\label{fig:271062192_large_separation}

\clearpage

\includegraphics[width=0.7\textwidth]{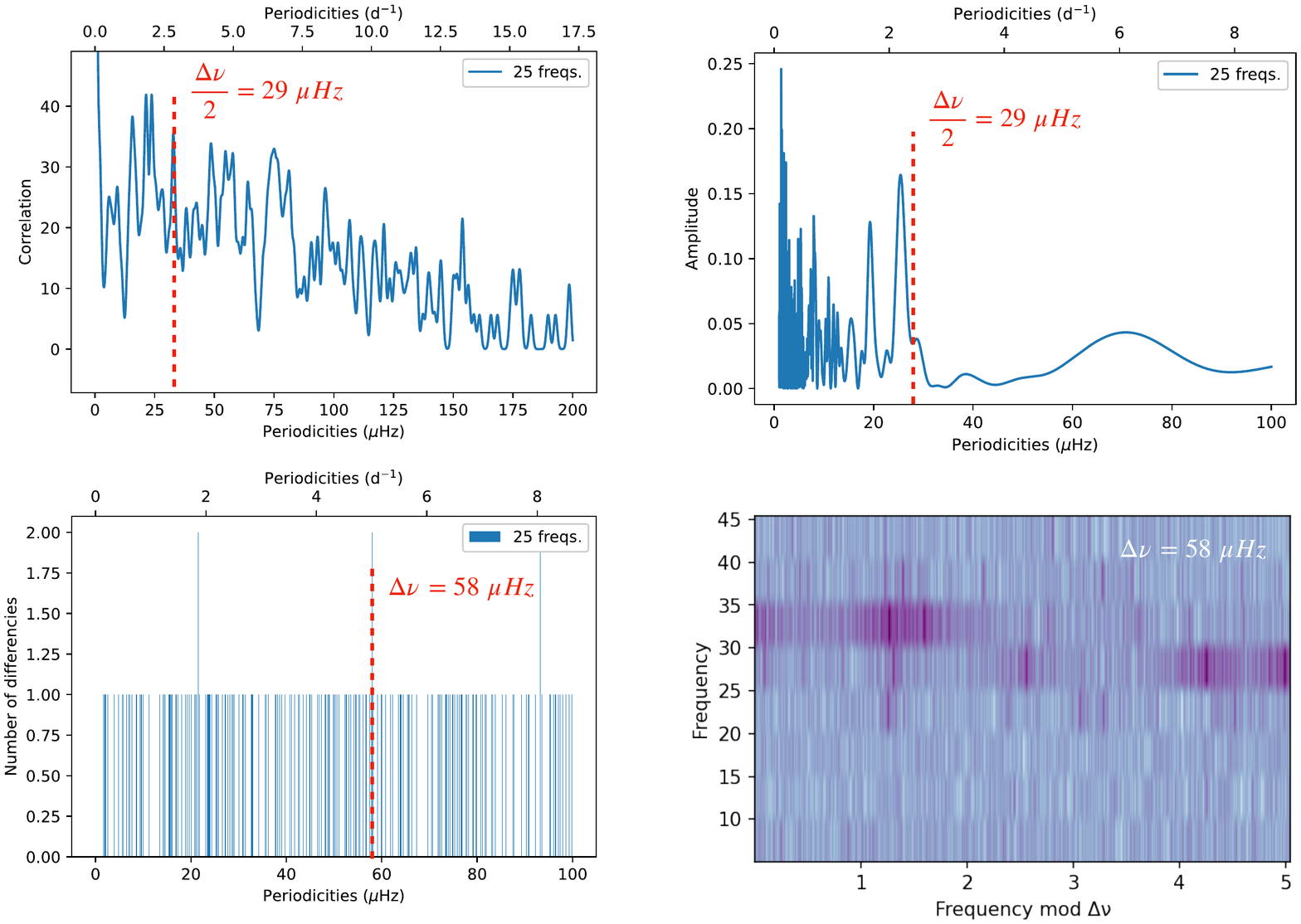}
\caption{The same as  Fig.~\ref{fig:TIC28943819_large_separation} for TIC\,175194881.}
\label{fig:175194881_large_separation}
\hfill \break
\hfill \break

\includegraphics[width=0.7\textwidth]{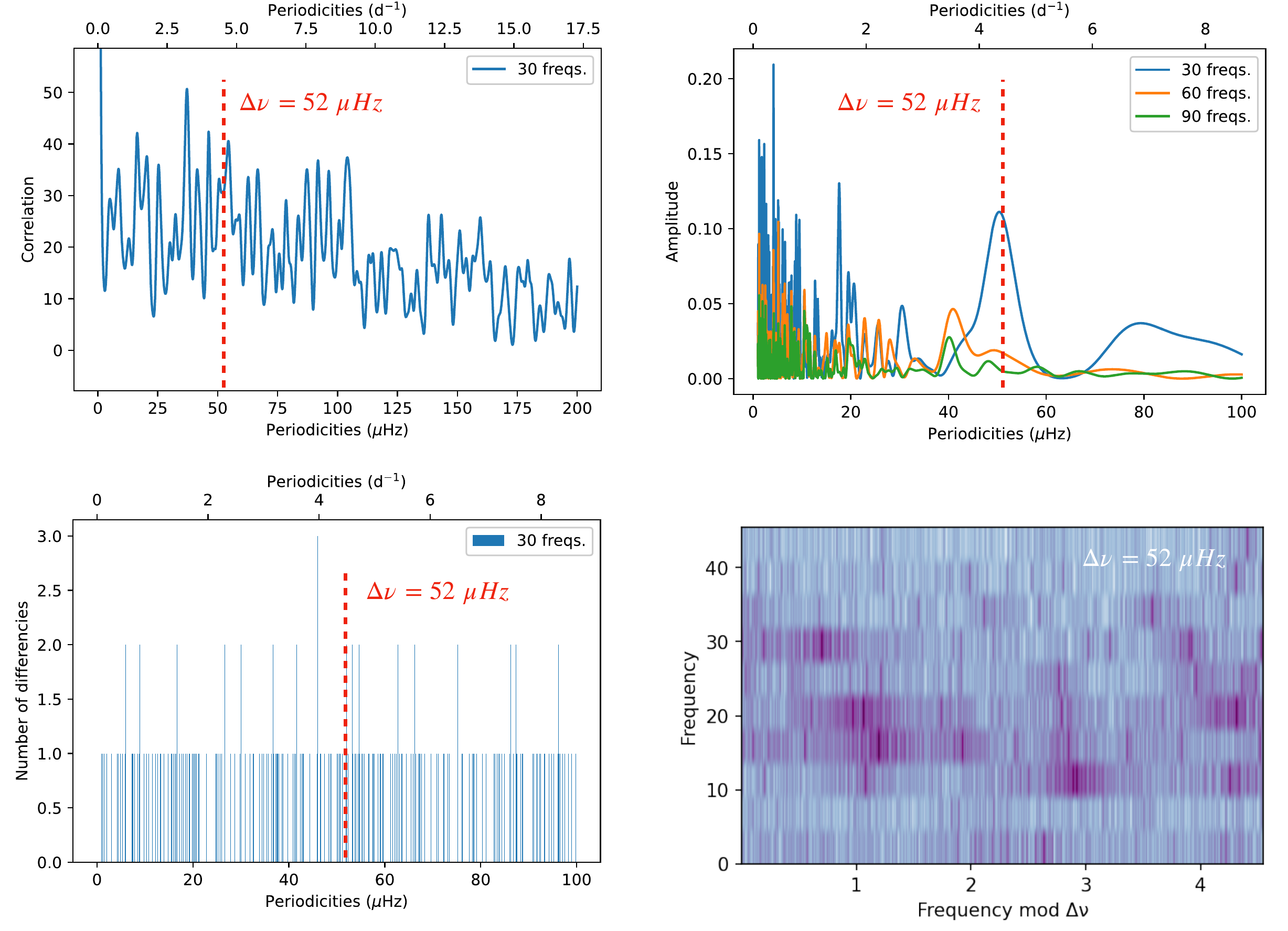}
\caption{The same as  Fig.~\ref{fig:TIC28943819_large_separation} for TIC\,175264376.}
\label{fig:175264376_large_separation}

\clearpage

\includegraphics[width=0.7\textwidth]{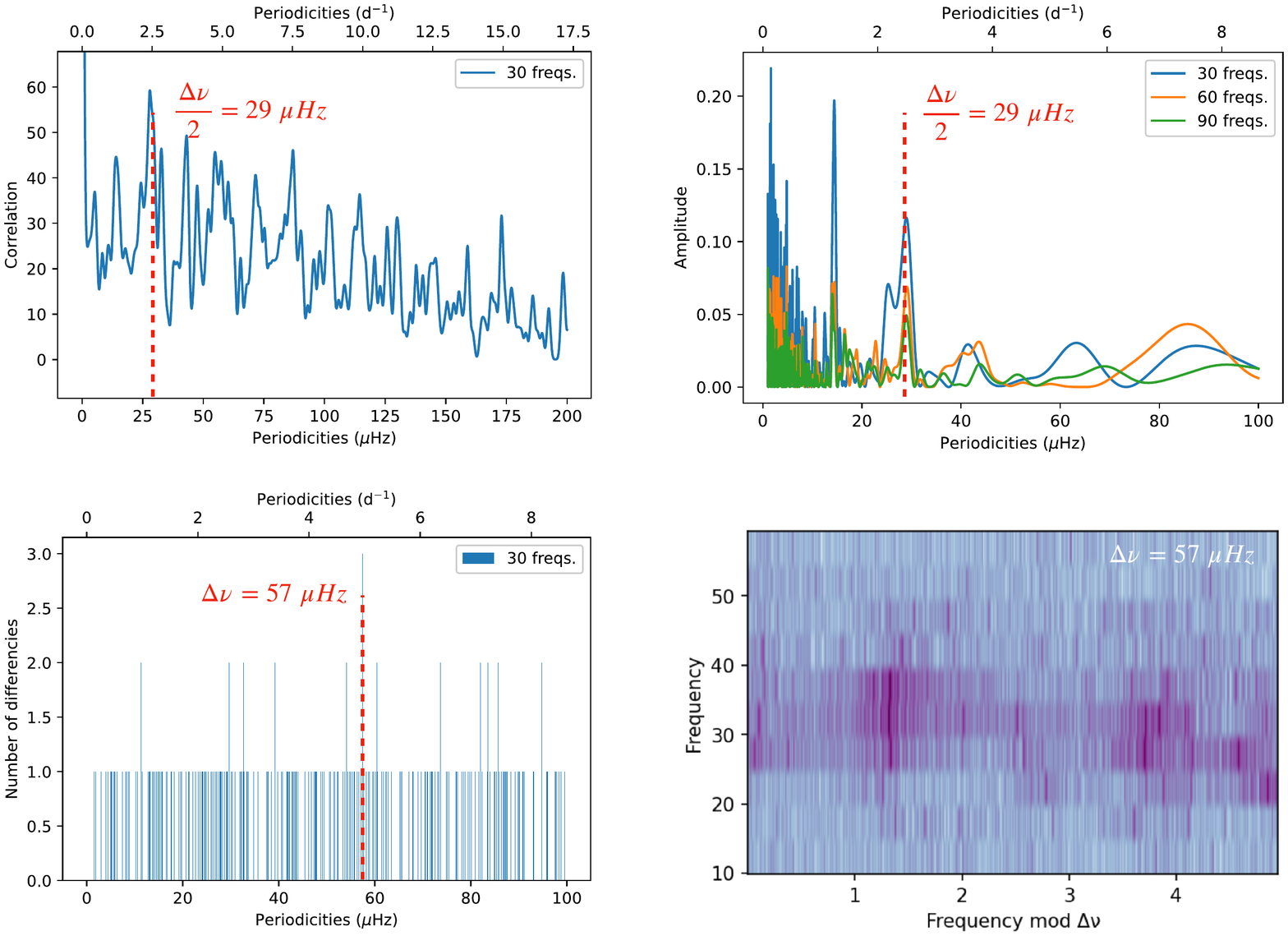}
\caption{The same as  Fig.~\ref{fig:TIC28943819_large_separation} for TIC\,175265807.}
\label{fig:175265807_large_separation}
\hfill \break
\hfill \break

\includegraphics[width=0.7\textwidth]{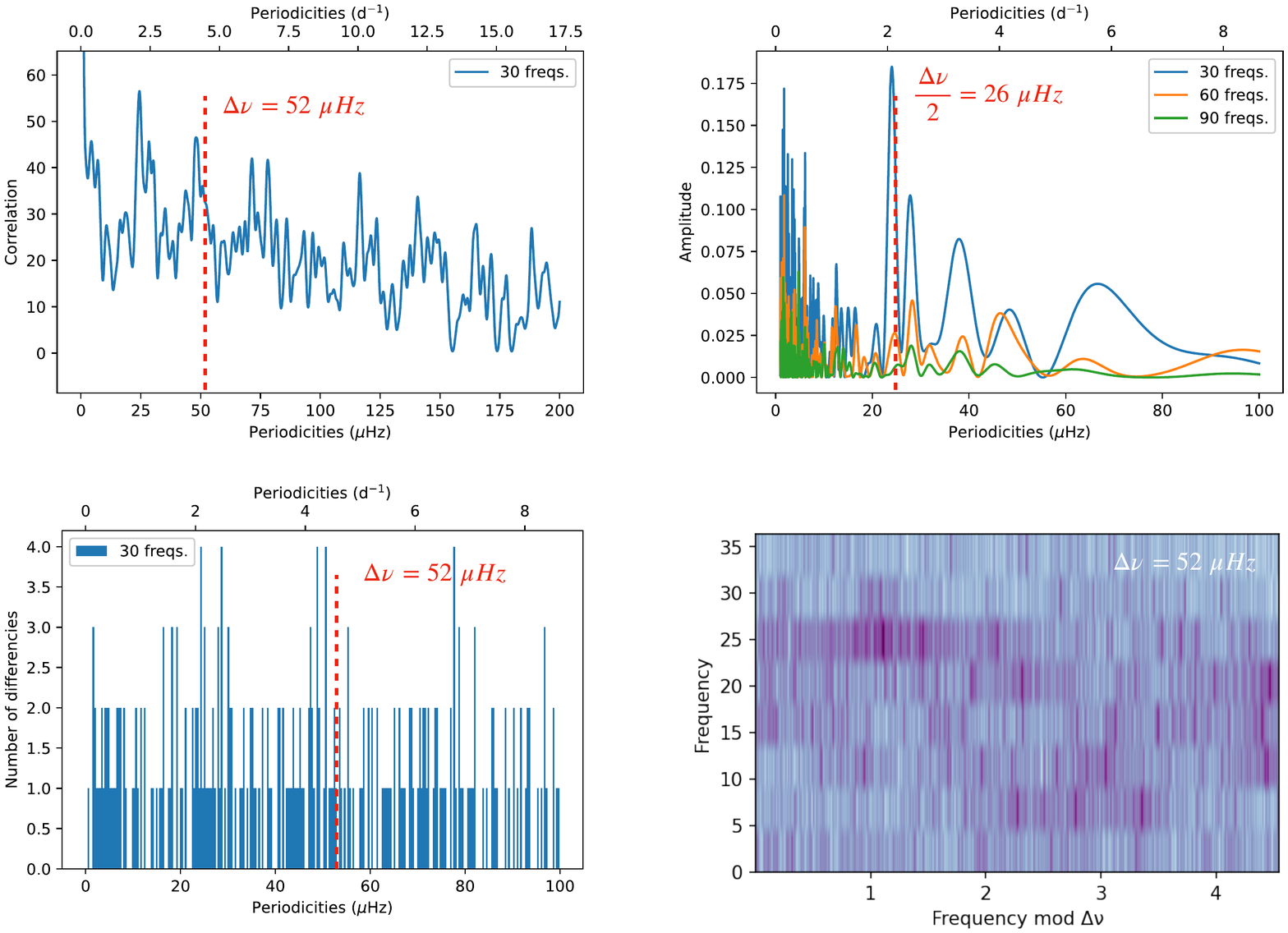}
\caption{The same as  Fig.~\ref{fig:TIC28943819_large_separation} for TIC\,175291778.}
\label{fig:175291778_large_separation}

\clearpage

\includegraphics[width=0.7\textwidth]{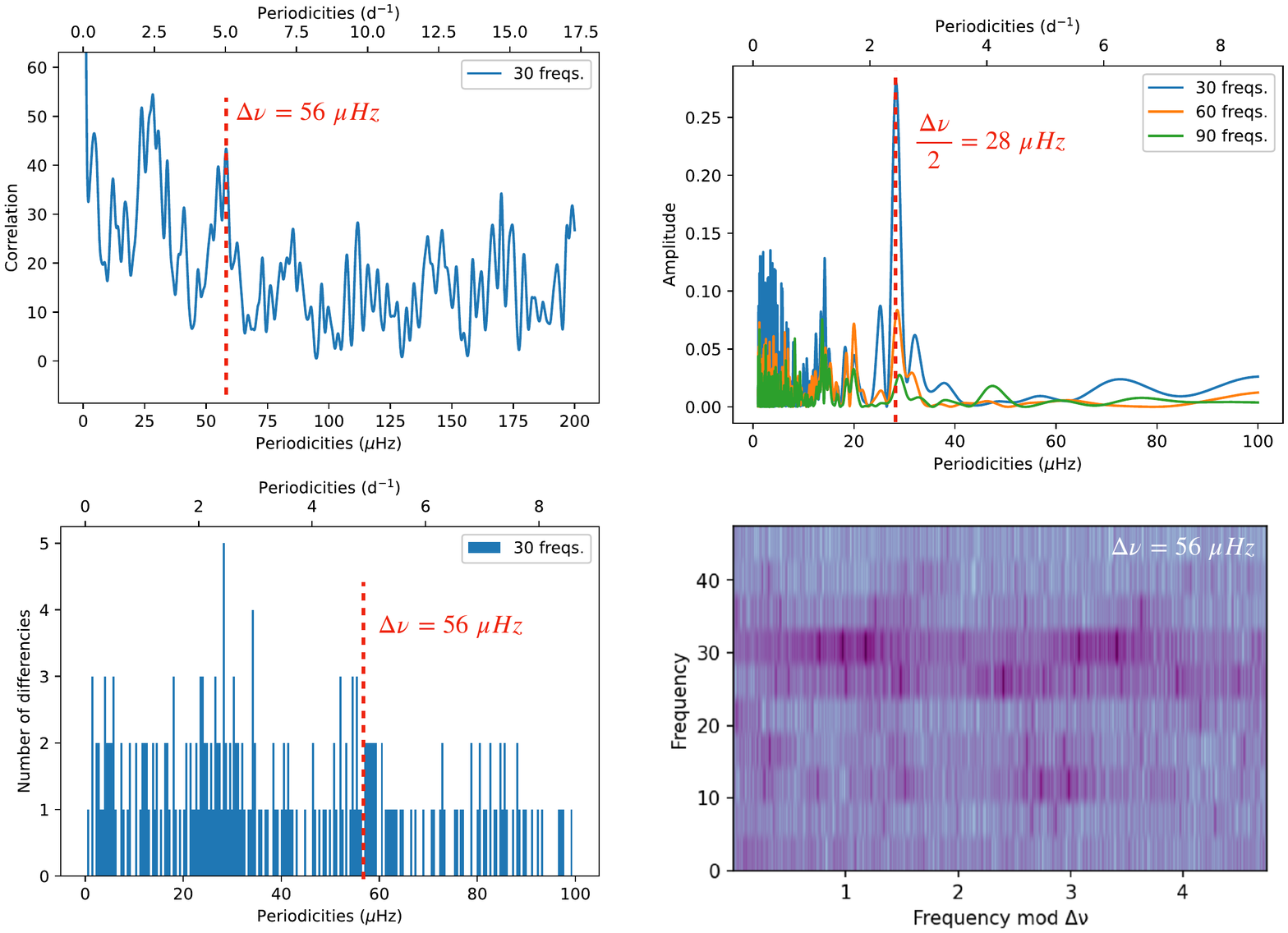}
\caption{The same as  Fig.~\ref{fig:TIC28943819_large_separation} for TIC\,184914505.}
\label{fig:184914505_large_separation}
\hfill \break
\hfill \break

\includegraphics[width=0.7\textwidth]{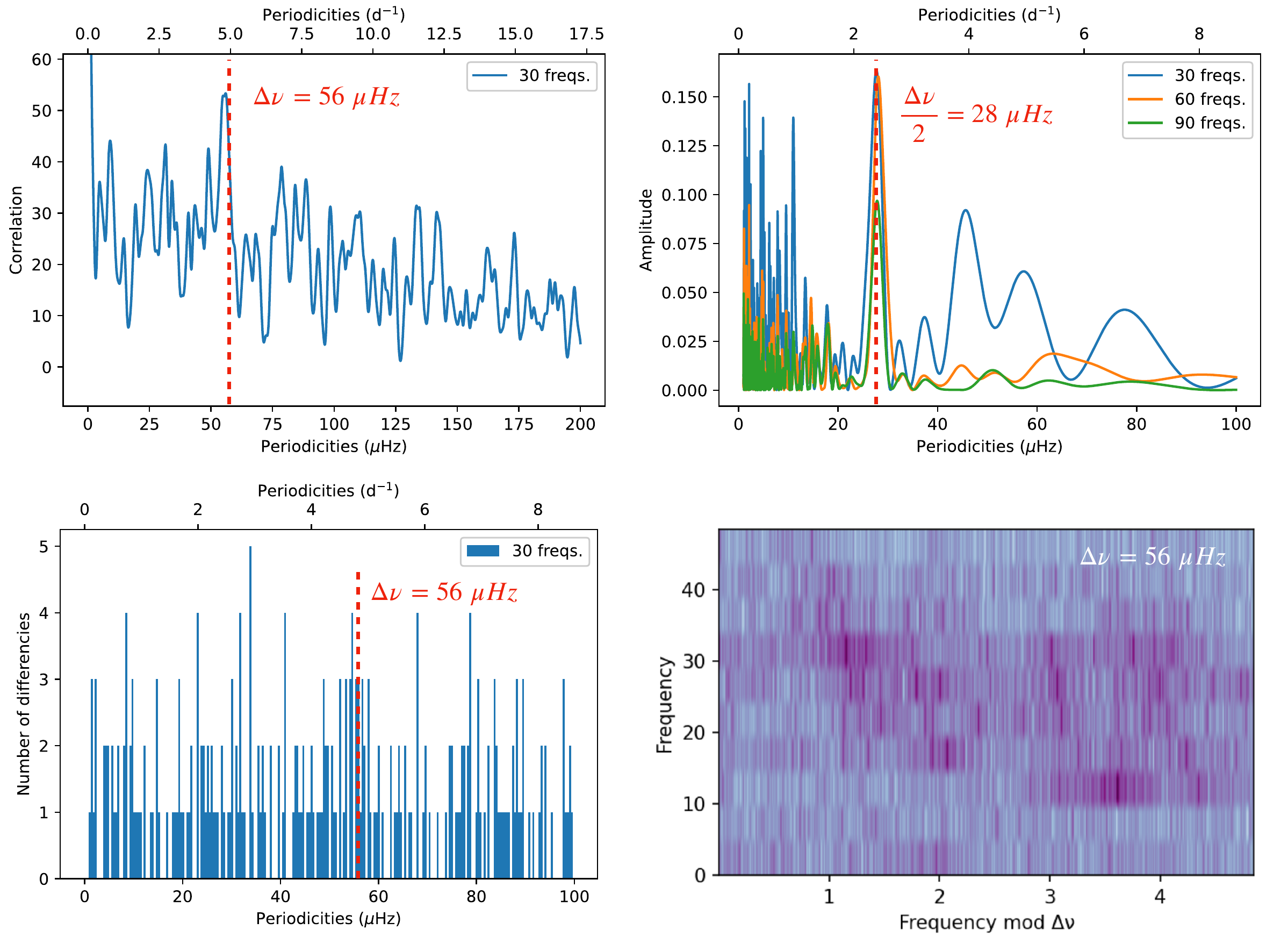}
\caption{The same as  Fig.~\ref{fig:TIC28943819_large_separation} for TIC\,184917633.}
\label{fig:184917633_large_separation}

\clearpage

\section{The positions and ranges of the possible radial modes of our $\delta$ Sct star sample}
\label{appendixC}
\setcounter{figure}{0} 
\renewcommand{\thefigure}{C.\arabic{figure}}
\hfill \break
\hfill \break
\includegraphics[width=0.7\textwidth]{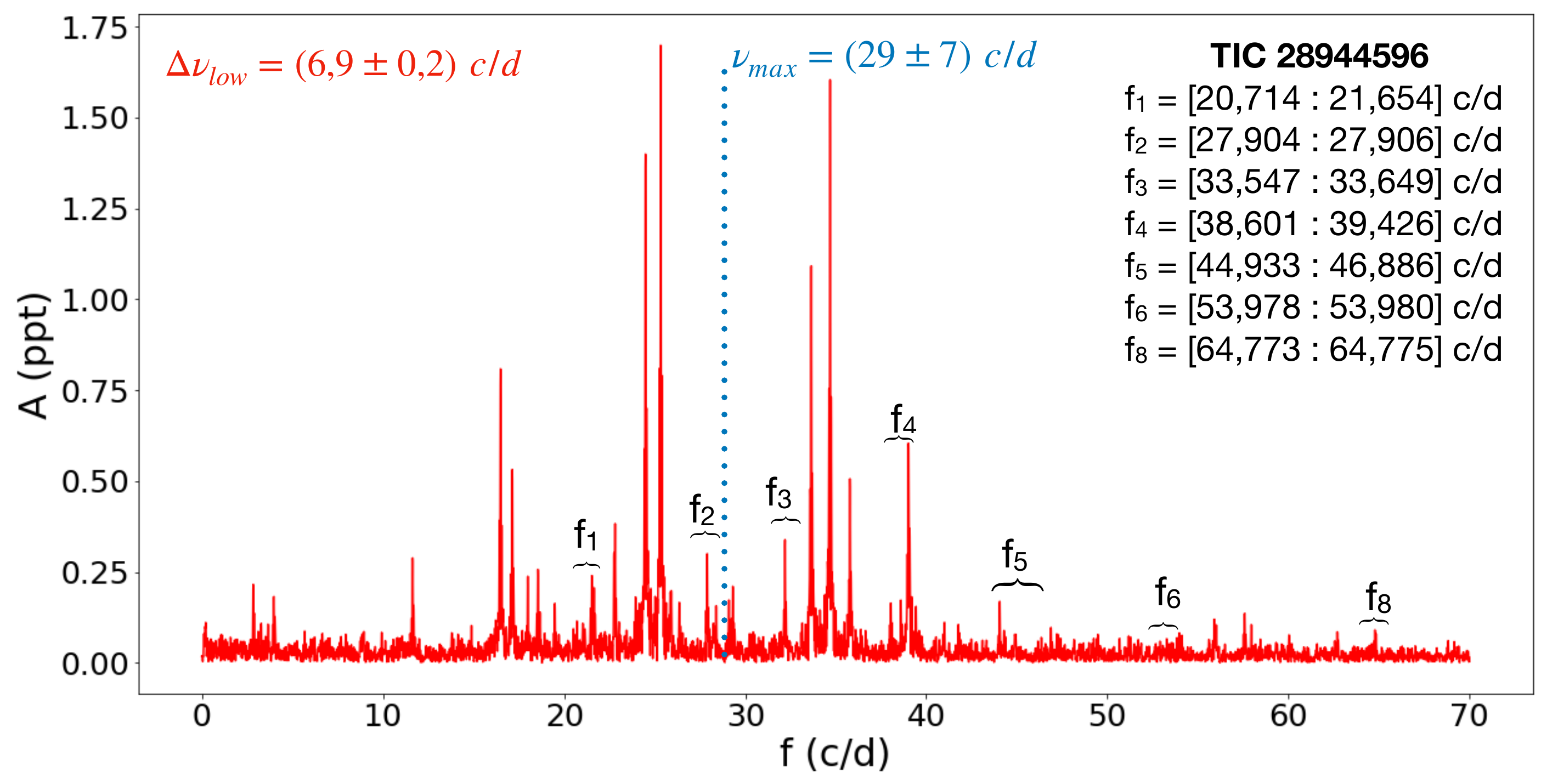}
\caption{The same as  Fig.~\ref{fig:TIC28943819_radial_overtones} for TIC\,28944596.}
\label{fig:28944596_radial_overtones}
\hfill \break
\hfill \break

\includegraphics[width=0.7\textwidth]{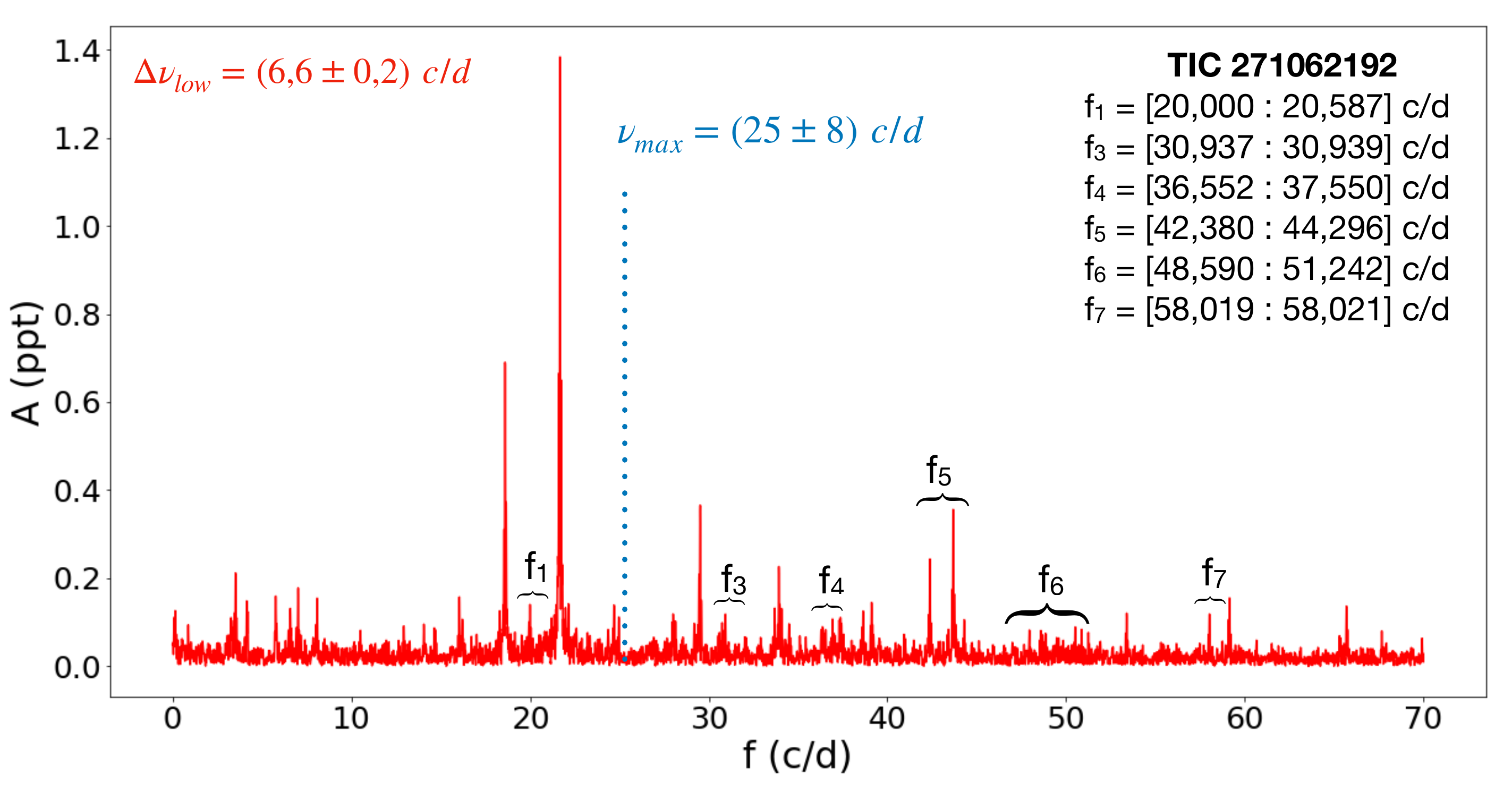}
\caption{The same as  Fig.~\ref{fig:TIC28943819_radial_overtones} for TIC\,271062192.}
\label{fig:271062192_radial_overtones}

\clearpage
\hfill \break
\hfill \break
\includegraphics[width=0.7\textwidth]{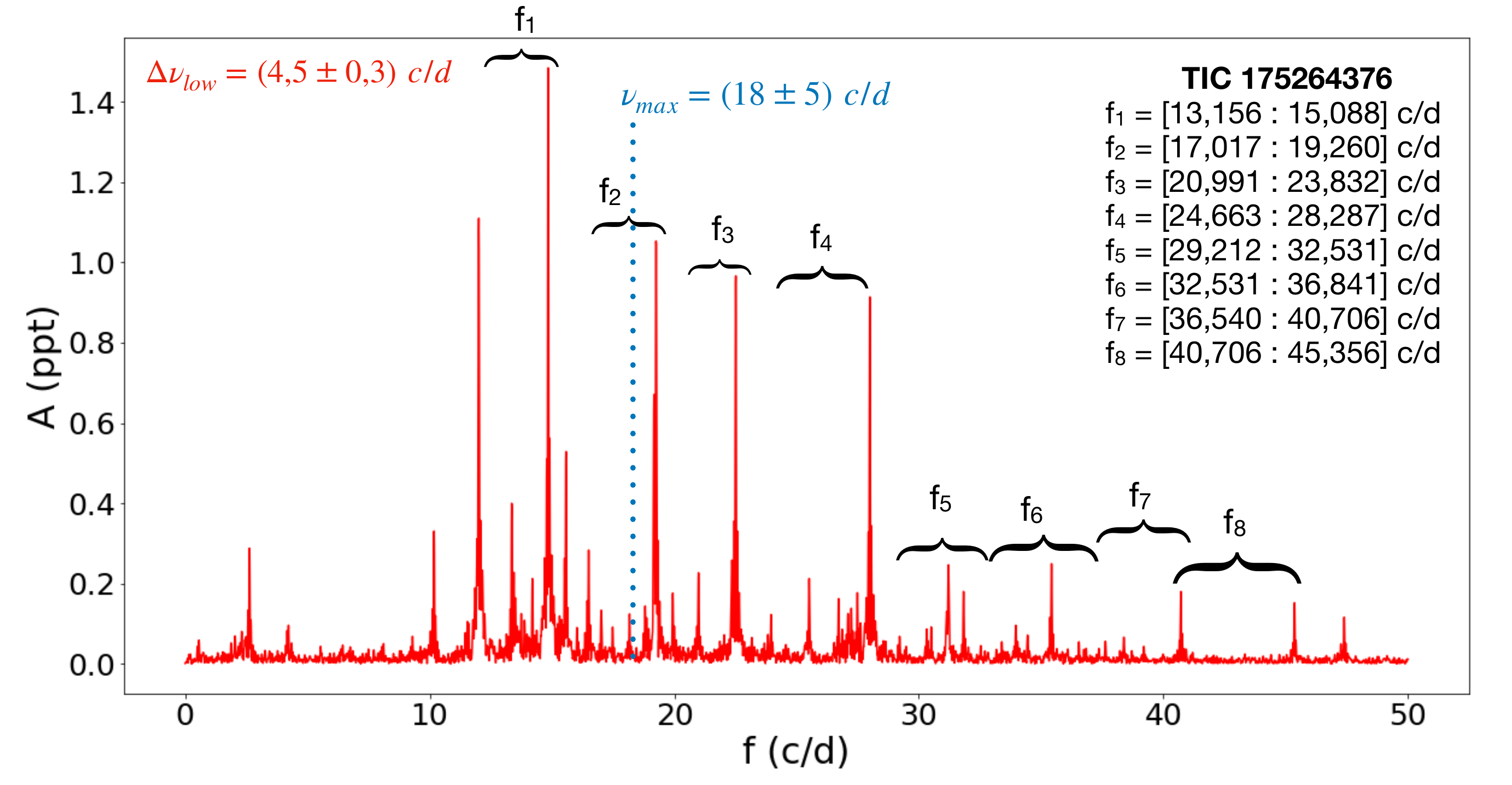}
\caption{The same as  Fig.~\ref{fig:TIC28943819_radial_overtones} for TIC\,175264376.}
\label{fig:175264376_radial_overtones}
\hfill \break
\hfill \break

\includegraphics[width=0.7\textwidth]{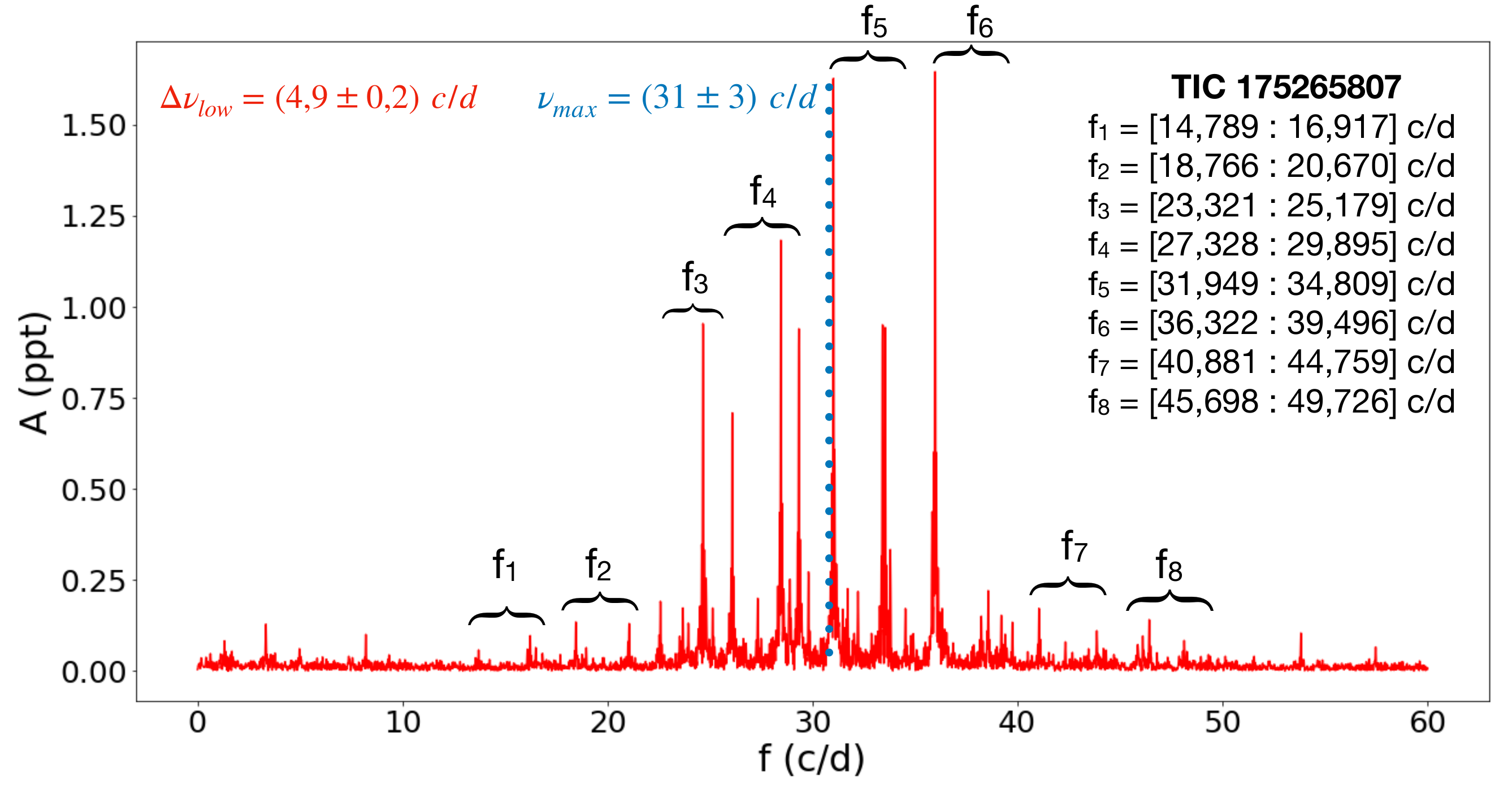}
\caption{The same as  Fig.~\ref{fig:TIC28943819_radial_overtones} for TIC\,175265807.}
\label{fig:175265807_radial_overtones}

\clearpage
\hfill \break
\hfill \break
\includegraphics[width=0.7\textwidth]{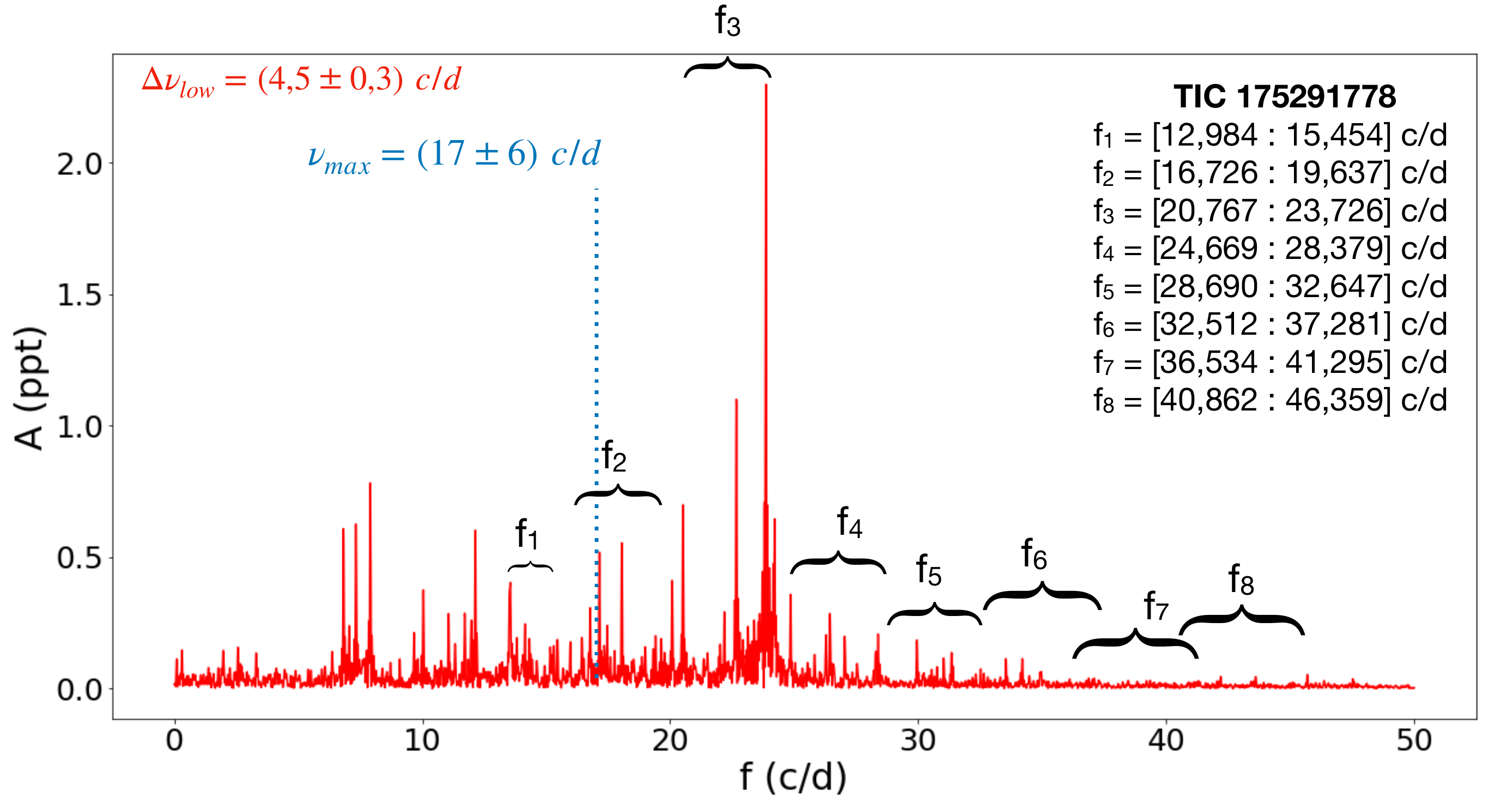}
\caption{The same as  Fig.~\ref{fig:TIC28943819_radial_overtones} for TIC\,175291778.}
\label{fig:175291778_radial_overtones}
\hfill \break
\hfill \break

\includegraphics[width=0.7\textwidth]{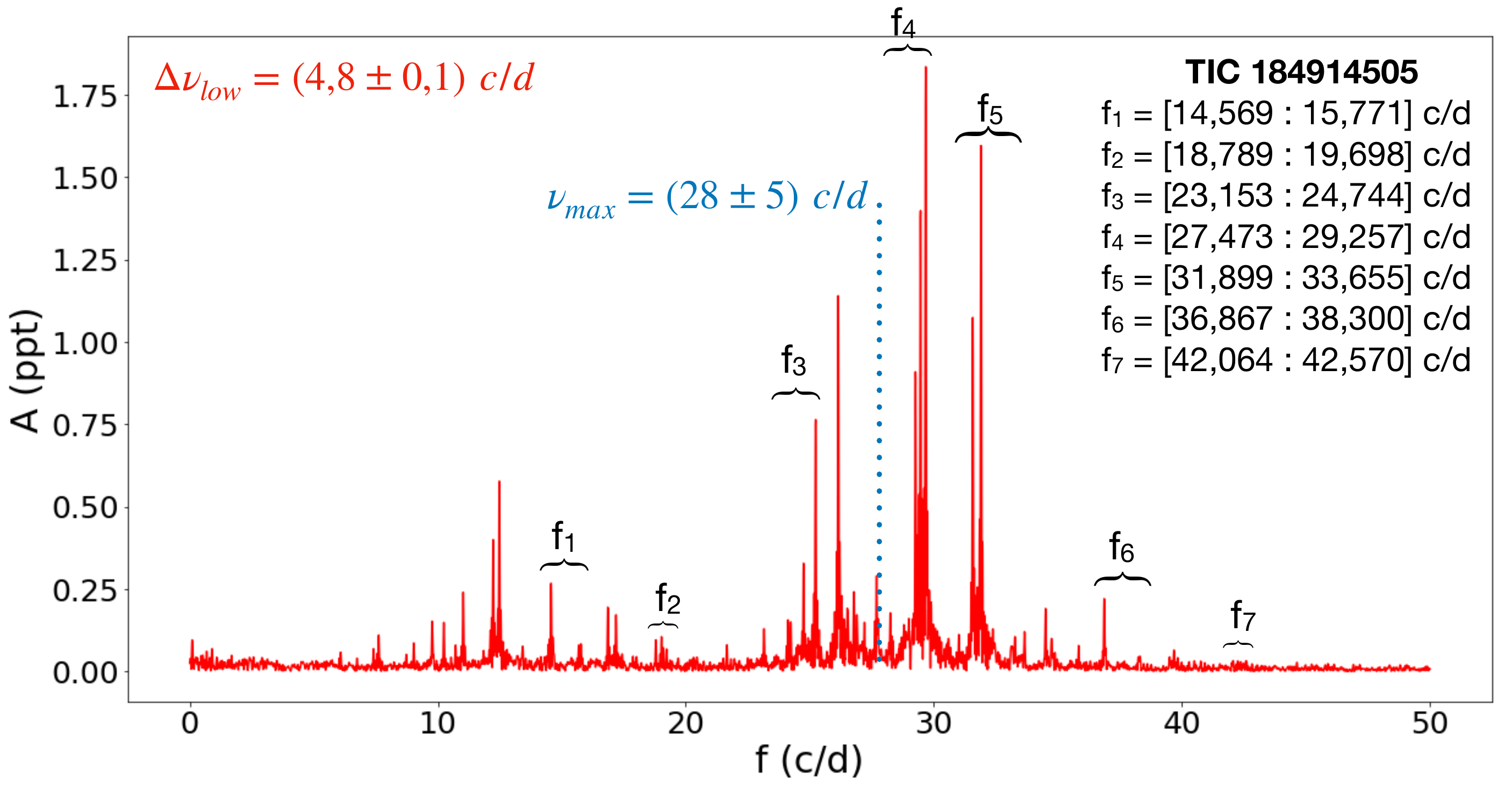}
\caption{The same as  Fig.~\ref{fig:TIC28943819_radial_overtones} for TIC\,184914505.}
\label{fig:184914505_radial_overtones}

\clearpage
\hfill \break
\hfill \break
\includegraphics[width=0.7\textwidth]{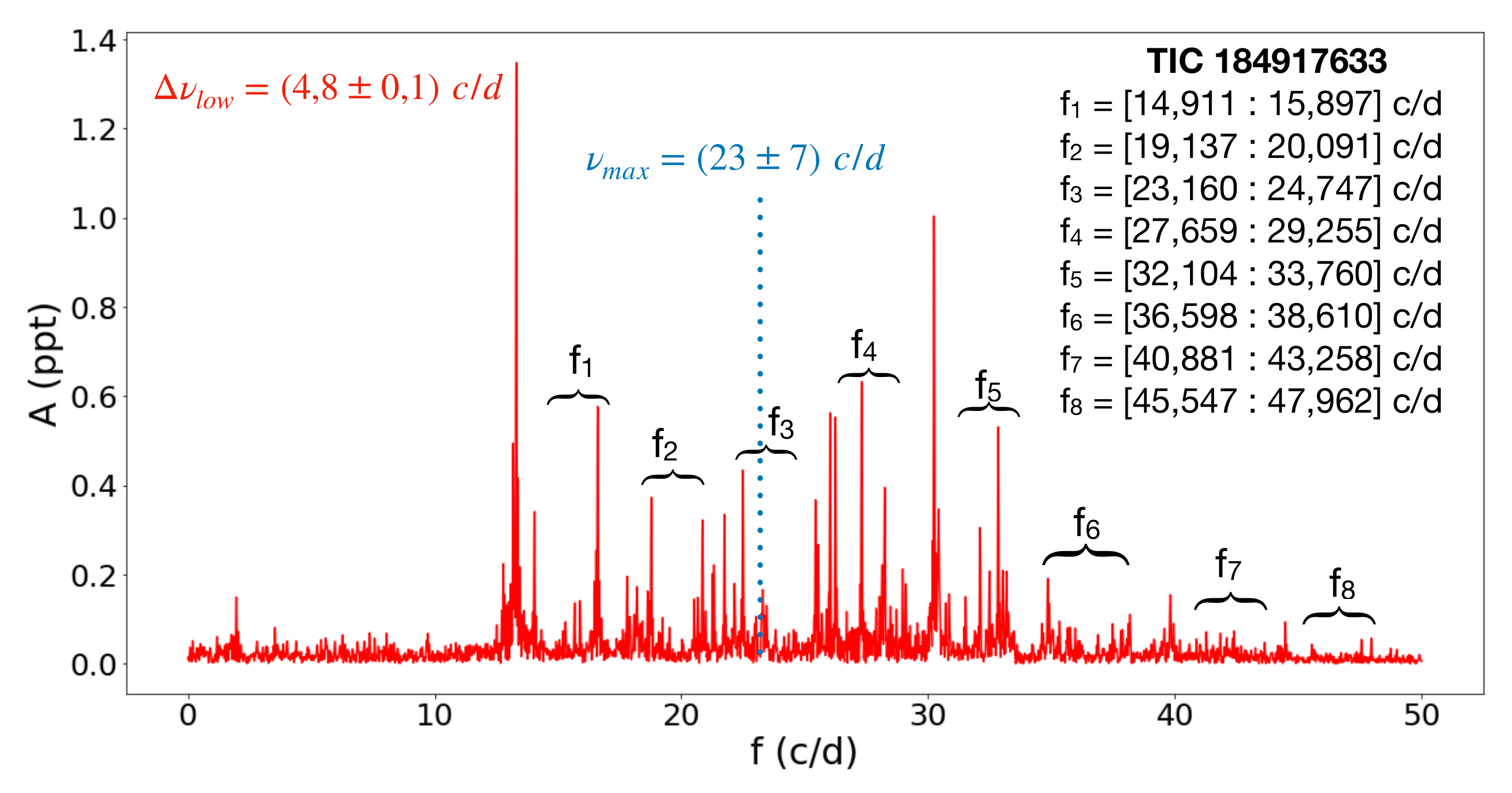}
\caption{The same as  Fig.~\ref{fig:TIC28943819_radial_overtones} for TIC\,184917633.}
\label{fig:184917633_radial_overtones}

\end{appendices}

\end{document}